\documentclass[sigconf,nonacm]{acmart}
\AtBeginDocument{%
  }

\usepackage{makecell}
\usepackage{array} 

\setcopyright{acmlicensed}
\copyrightyear{2018}
\acmYear{2018}
\acmDOI{XXXXXXX.XXXXXXX}
\acmConference[Conference acronym 'XX]{Make sure to enter the correct
  conference title from your rights confirmation email}{June 03--05,
  2018}{Woodstock, NY}
\acmISBN{978-1-4503-XXXX-X/2018/06}

\begin{document}

\title{MazeMate: An LLM-Powered Chatbot to Support Computational Thinking in Gamified Programming Learning}


\author{Chenyu Hou}
\authornote{Both authors contributed equally to this research.}
\email{chenyu004@e.ntu.edu.sg}
\affiliation{%
  \institution{Nanyang Technological University}
  \country{Singapore}
}

\author{Hua Yu}
\authornotemark[1]
\email{yu_hua@ntu.edu.sg}
\affiliation{%
  \institution{Nanyang Technological University}
  \country{Singapore}
}

\author{Gaoxia Zhu}
\email{gaoxia.zhu@nie.edu.sg}
\affiliation{%
  \institution{National Institute of Education, Nanyang Technological University}
  \country{Singapore}}

\author{John Derek Anas}
\email{johnderek.anas@ntu.edu.sg}
\affiliation{%
  \institution{Nanyang Technological University}
  \country{Singapore}
}

\author{Jiao Liu}
\email {jiao.liu@ntu.edu.sg}
\affiliation{%
  \institution{Nanyang Technological University}
  \country{Singapore}
}

\author{Yew Soon Ong}
\email{ASYSOng@ntu.edu.sg}
\affiliation{%
  \institution{Nanyang Technological University}
  \country{Singapore}
}
\begin{abstract}
Computational Thinking (CT) is a foundational problem-solving skill, and gamified programming environments are a widely adopted approach to cultivating it. While large language models (LLMs) provide on-demand programming support, current applications rarely foster CT development. We present MazeMate, an LLM-powered chatbot embedded in a 3D Maze programming game, designed to deliver adaptive, context-sensitive scaffolds aligned with CT processes in maze solving and maze design. We report on the first classroom implementation with 247 undergraduates. Students rated MazeMate as moderately helpful, with higher perceived usefulness for maze solving than for maze design. Thematic analysis confirmed support for CT processes such as decomposition, abstraction, and algorithmic thinking, while also revealing limitations in supporting maze design, including mismatched suggestions and fabricated algorithmic solutions. These findings demonstrate the potential of LLM-based scaffolding to support CT and underscore directions for design refinement to enhance MazeMate’s usability in authentic classrooms.
\end{abstract}



\received{20 February 2007}
\received[revised]{12 March 2009}
\received[accepted]{5 June 2009}

\maketitle

\section{Introduction}
Computational Thinking (CT) has emerged as a critical 21st-century problem-solving skill, broadly defined as the thinking processes used to formulate problems and represent their solutions that enable an information-processing agent to execute them effectively~\cite{wing2006computational}.
One widely adopted approach to cultivating CT is through gamified programming environments~\cite{cheng2017teaching}. 
There are multiple gamified platforms such as Scratch \cite{moreno2016code}, Alice \cite{costa2017relation}, and Code.org \cite{kaleliouglu2015new} that situate programming within interactive visual challenges that make CT concepts accessible to novices while fostering engagement and creativity ~\cite{weintrop2018starting,kazimoglu2020enhancing,brennan2012new}. For example, Scratch is a widely-used platform that allows users to create interactive stories, games, and animations with block-based code \cite{chang2014effects,weintrop2018starting}. Similarly, 3D Maze is a block-based programming game platform that integrates visual design with algorithms. This platform has been used in various educational contexts, such as digital literacy and engineering design \cite{zhu2025exploring,zhu2024human}. 
In 3D Maze, the maze solving mode emphasizes logical rigor and algorithmic reasoning by guiding students to express movement concepts using code blocks (e.g., loops, conditionals). In contrast, the maze design mode externalizes reasoning through visual and narrative elements, encouraging students to connect design patterns with computational constructs such as loops and conditional logic~\cite{basawapatna2013simulation,banic2019visual}. Together, solving and designing activities strengthen both the technical and creative dimensions of CT \cite{wu2025integrating}.

While gamified environments provide an accessible entry point into programming, research demonstrates that scaffolding is essential for cultivating computational thinking effectively~\cite{tikva2023effect}. 
Traditional scaffolding approaches in gamified environments can support novices, but they are often limited in flexibility and cannot dynamically adapt to students’ evolving problem states~\cite{wang2025scaffolding,ma2025dbox}. The emergence of generative AI \cite{thirunavukarasu2023large} introduces new possibilities for scaffolding that is adaptive, contextualized, and responsive. Large Language Models (LLMs) can tailor support to students’ needs by adjusting explanations, generating examples and responding to evolving problem contexts \cite{kasneci2023chatgpt}. 
For example, one study~\cite{ma2025dbox} demonstrated that DBox, an LLM-based chatbot for programming education, effectively scaffolded problem decomposition and improved students’ agency in programming tasks. However, their work primarily targeted a single stage of CT and did not fully capture the complete computational thinking process that spans decomposition, pattern recognition, abstraction, and algorithmic thinking. At present, there is still a lack of LLM-based scaffolding designs that support the entire CT learning process.


To address these challenges, this paper presents MazeMate, an LLM-powered chatbot embedded in the 3D Maze environment to provide pedagogical scaffolds for fostering computational thinking.
MazeMate is designed to provide pedagogically grounded scaffolding that aligns AI support with the stages of CT in both maze design and maze solving activities. 
In maze design activity, MazeMate guides students to break down design requirements, recognize structural patterns, and abstract layout rules into solvable configurations. In maze solving activity, which is inherently a complex and logic-intensive task involving multiple monsters, gems, and resource constraints, MazeMate supports students in transforming step-by-step actions into algorithmic solutions. 
Unlike general-purpose LLMs that often hallucinate in high-complexity and logic settings \cite{chu2025llm}, such as the maze solving task, MazeMate mitigates this risk by synchronizing with real-time maze data and integrating algorithmic backends for solution verification, ensuring that scaffolds remain accurate in logic task. In summary, MazeMate has the four design features: contextual awareness of the 3D Maze environment, adaptive support based on real-time progress, integration of correct solutions as contextual knowledge, and delayed reveal of answers to prevent cognitive shortcuts. For classroom implementation of MazeMate, these design features were further aligned with classroom lesson plans to ensure that MazeMate supports computational thinking in authentic educational contexts.

In the following sections, we first introduce the background 3D Maze environment, i.e., a platform designed to support collaborative programming and computational thinking across diverse student groups (Section \ref{Section 3}).
To identify the limitations of general-purpose LLMs and clarify MazeMate’s design goals, we conducted a formative study during an earlier round of 3D Maze implementation and analyzed students’ reflection responses. A thematic analysis revealed four key design goals (Section \ref{Section 3.2}), which informed MazeMate’s development. We then detail MazeMate’s design features, including path-finding algorithms to address LLM struggles with complex mazes, and the alignment of scaffolding strategies with core computational thinking processes (Section \ref{Section 4}). Finally, we report findings from a user study with 247 undergraduate students. Their survey responses provide insights into the effectiveness of MazeMate’s design and inform implications for further refinement (Section \ref{Section 5}). The contributions of this paper can be summarized as follows:
\begin{itemize}
    \item We present \textbf{MazeMate}, a LLM–powered chatbot embedded in the 3D Maze programming game, designed to scaffold CT in authentic classrooms.
    \item We provide two \textbf{design structures }that align MazeMate scaffolds with the core CT processes (i.e., decomposition, pattern recognition, abstraction, and algorithmic thinking) across maze solving and maze design activities.
    \item We provide \textbf{empirical insights} from the first classroom implementation with 247 undergraduates, highlighting how students perceived and engaged with MazeMate to support CT.
    \item We outline \textbf{design implications} for developing and refining LLM-based pedagogical chatbots to better support CT development.
\end{itemize}


\begin{figure*}
    \centering
    \includegraphics[width=1\linewidth]{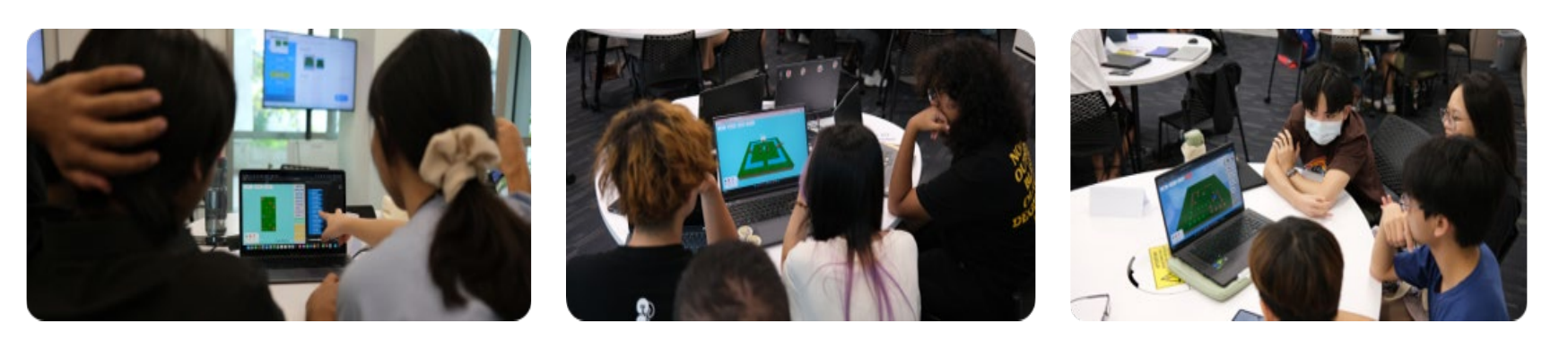}
    \caption{Collaborative with group members in the 3D maze system.}
    \label{fig:Collaborate}
\end{figure*}

\section{Related Work}
\subsection{Computational Thinking and Programming Learning}
Computational Thinking (CT) is an analytical skill for problem-solving~\cite{wing2006computational}. Despite differences across theoretical models, there is growing consensus that CT comprises four interrelated components: decomposition, pattern recognition, abstraction, and algorithmic thinking~\cite{selby2013computational}. Decomposition involves breaking complex problems into manageable parts~\cite{lye2014review}; pattern recognition and abstraction highlight recurring structures, simplifying complexity by focusing on essential features that could be automated~\cite{wing2006computational}; and algorithmic thinking refers to constructing step-by-step procedures for solving problems~\cite{shute2017demystifying}. CT is a foundational skill for all, not only programmers, as it equips individuals to manage information effectively in today’s data-driven era~\cite{sanford2016computational,wing2017computational,burke2016computational}. 

At its essence, computational thinking is a systematic approach to tackling complex issues, and therefore, it is indispensable for students from diverse backgrounds \cite{wing2006computational}. Owing to the abstract nature of CT~\cite{hsu2025programming}, gamified programming environments have been increasingly used, which allow students with varying levels of programming expertise to engage with CT processes in an interactive and engaging way~\cite{hsu2018learn}. For example, one study investigated students’ experiences with visual programming tools such as Alice and Scratch. Through a semester-long study where participants created visual projects using block-based code, they found that such environments enhanced learning engagement in introductory programming courses~\cite{chang2014effects}. In terms of the effect of a gamified programming environment on CT, one study analyzed Scratch game projects and found that such game-based activities naturally cultivate CT, such as decomposition and algorithm thinking~\cite{hoover2016assessing}. On the other hand, another study took a more process-oriented assessment of CT development, as 8th-grade students designing games in Scratch. They found that game design facilitated growth in logical reasoning and synchronization, but more abstract skills such as decomposition were harder to develop~\cite{troiano2019my}. These findings suggest that, while gamified programming environments can promote aspects of CT, they may not consistently cultivate all components. Consequently, it is vital to integrate additional support in the gamified programming environment to ensure that students practice the full processes of computational thinking~\cite{hsu2018learn}. 

\subsection{Scaffolding for Computational Thinking}
In educational psychology, scaffolding refers to ``a process that enables a child or novice to solve a task or achieve a goal that would be beyond their unassisted efforts” \cite{wood1976role}. Research has been implementing scaffolding in programming games to foster computational thinking. For example, one study designed three scaffolding techniques in a maze-based programming game, including semi-finished solutions, explicit instructions on CT concepts, and explanations for the logic of solution design. They found that, compared to the non-scaffolding group, the scaffolding group significantly improved the CT scores, measured by a self-report questionnaire~\cite{tikva2023effect}. Similarly, another study conducted a quasi-experiment to compare four conditions of scaffolding in fostering CT, including guidance (i.e., a worksheet that provides metacognitive questions), no guidance, immediate feedback (i.e., a visual feedback of implementing the code), and delayed feedback (a visual feedback delayed 30 seconds after implementing the code). They found that delayed feedback is an effective scaffolding strategy that fosters the algorithm reasoning of CT~\cite{chevalier2022role}. 

The emergence of Generative AI has opened up new venues for supporting programming learning \cite{prather2023s}. Large language models (LLMs) such as ChatGPT and Codex can serve as on-demand coding assistants, helping with tasks such as code completion, debugging, translation, summarization, and explanation of code~\cite{kazemitabaar2023novices}. However, recent reviews of integrating LLMs into programming education caution against overreliance, as students might copy AI-generated solutions and bypass the computational thinking process \cite{raihan2025large,cambaz2024use,pirzado2024navigating}.
In addition, researchers reported that current LLM agent for education still faces challenges in handling complex tasks, such as programming and path planning tasks \cite{angeli2020computational,triantafyllou2024gamification,chu2025llm}.

Instead, LLMs should be involved in a gamified programming environment as scaffolding tools to foster computational thinking~\cite{yan2025llm}. Research has designed different systems to carefully scaffold students' practice of computational thinking, where the student remains actively involved, guiding the AI in smaller steps or integrating AI suggestions with their own work, which can yield better learning gains. For example, CodeTailor, a LLM-powered personalized agent, can tailor several hints based on students' latest code, so as to help students identify code errors and improve their code \cite{hou2024codetailor}.  More closely aligned with computational thinking, DBox is an LLM-powered chatbot that supports students in decomposing algorithmic problems by assisting in constructing students' coding ideas and providing progressive hints for correct code \cite{ma2025dbox}. The findings demonstrated that, compared to an unstructured AI helper, the scaffolded DBox tool significantly improved students’ problem-solving abilities and even their confidence and sense of ownership over the work \cite{ma2025dbox}. This prior work has shed light on the potential of utilizing LLM in scaffolding computational thinking. However, these works are limited as they do not capture the complete process of computational thinking in solving complex issues. Thus, our work built on these insights to design a novel LLM-based chatbot embedded in a gamified programming environment to support the four core stages of computational thinking (i.e., decomposition, pattern recognition, abstraction, and algorithmic thinking).

\section{BACKGROUND: 3D MAZE} \label{Section 3}
\subsection{3D Maze Introduction and Activities}
The 3D Maze system is an interactive learning tool for programming learning developed on the Unity platform, integrating two major modules: maze design and maze solving. As shown in Fig.~\ref{fig:Collaborate}, in this collaborative programming game, group members can take on distinct roles in maze design and maze solving, and display collective responsibility. In the following, we will introduce the two activities in detail.

\begin{figure*}
    \centering
    \includegraphics[width=1\linewidth]{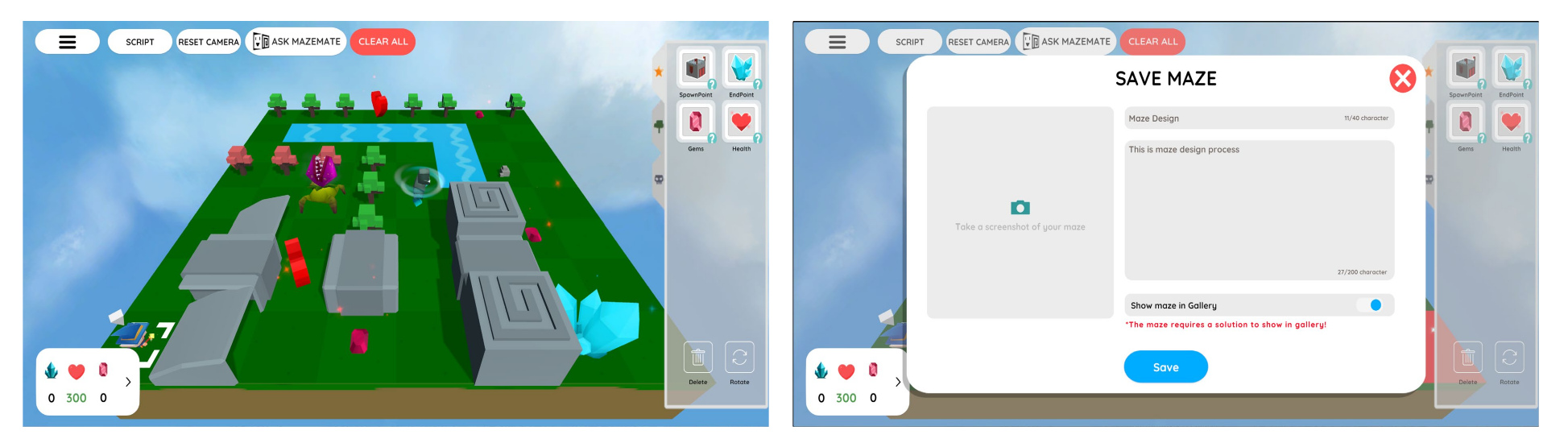}
    \caption{Maze design process in the 3D maze system.}
    \label{fig:3DMazeDesign}
\end{figure*}

\textbf{Maze design}.
3D maze design is an open-ended, creativity-centered activity. In the maze design activity, students can design mazes with diverse and novel layouts. In the 3D Maze environment, we incorporated various environmental assets, e.g., gems, hearts, different obstacles, and different monsters. In our platform, players must collect all gems before reaching the goal, while engaging with enemies that block the path. Each enemy type inflicts a fixed amount of damage, requiring the player’s Avatar to spend health points to defeat them, for example, Bat (–20), Ghost (–40), Skeleton Archer (–20), and Dragon (–60). To succeed, the total health consumed across mandatory encounters must remain lower than the Avatar’s initial and collected health. This rule introduces a resource-management layer, requiring players to plan routes and prioritize battles strategically. As shown in Fig.~\ref{fig:3DMazeDesign}, after completing a design, students can publish their maze to a shared Gallery, making it discoverable and playable by peers who can attempt solutions.


\textbf{Maze solving}.
3D maze solving is a logic-intensive, constraint-rich programming task; students must translate paths into generalizable algorithms with conditionals and loops.
During the maze solving activity, the 3D Maze visualizes the solution process, enabling students to test different path strategies while exercising algorithmic thinking. Instead of enumerating step-by-step moves, students are required to express solutions using logic-like statements with condition blocks (e.g., if-else condition) and iterative control (e.g., while loop). Only such representations count as \textbf{High-Efficiency Solutions}, whereas simple action sequences are classified as \textbf{Low-Efficiency Solutions} and do not meet the task requirement. The differences between these two kinds of solutions are presented in Table~\ref{fig:3dprint}. This design shifts the emphasis from reproducing a path to articulating an underlying algorithm, thereby aligning the activity with computational thinking.


\begin{table*}[!t]
  \centering
  \begin{tabular}{c|c|c}
    \hline
     Maze & Low Efficiency Solution & High Efficiency Solution \\
    \hline
     \makecell{\includegraphics[width=0.31\textwidth]{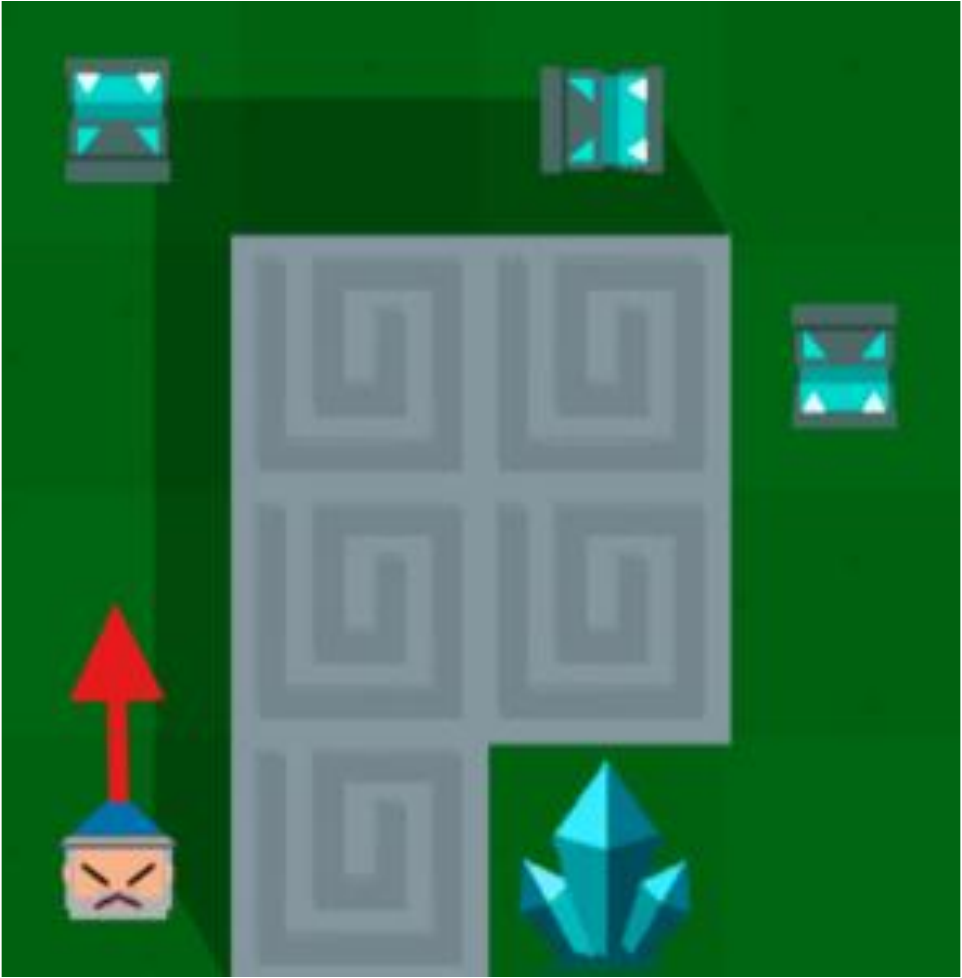}} &
     \makecell{\includegraphics[width=0.28\textwidth]{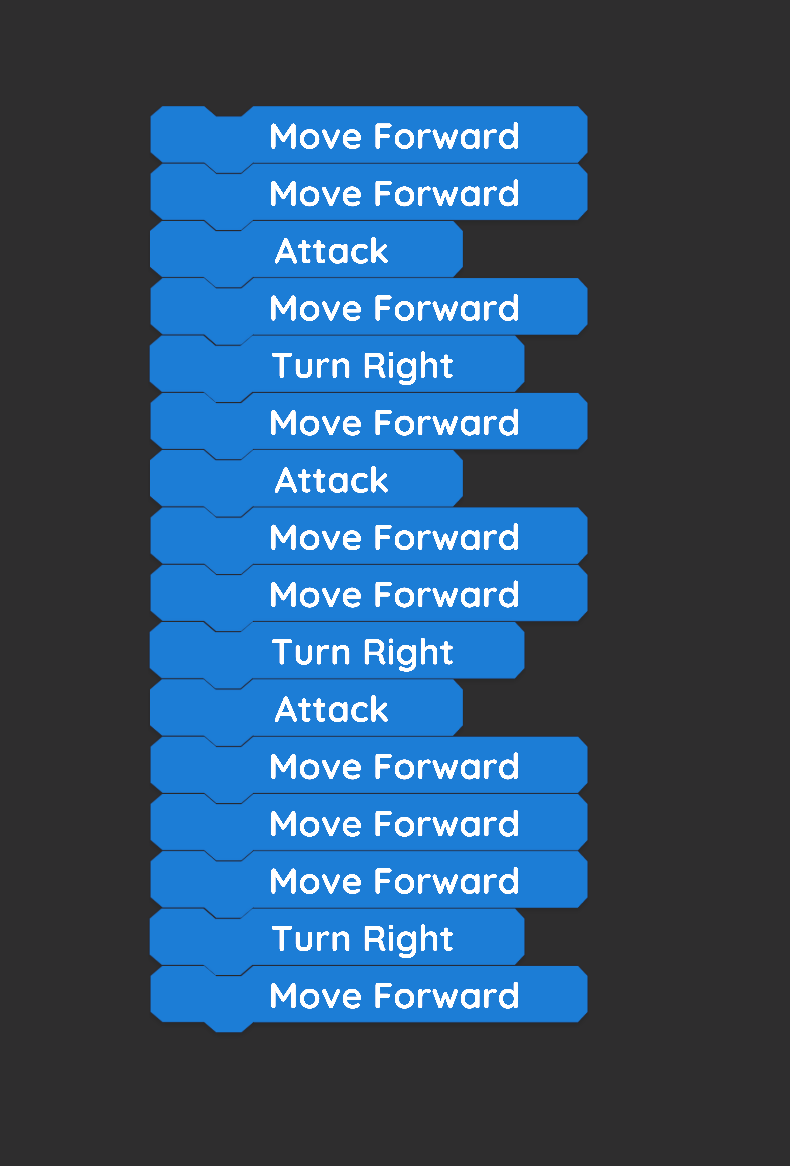}} &
     \makecell{\includegraphics[width=0.28\textwidth]{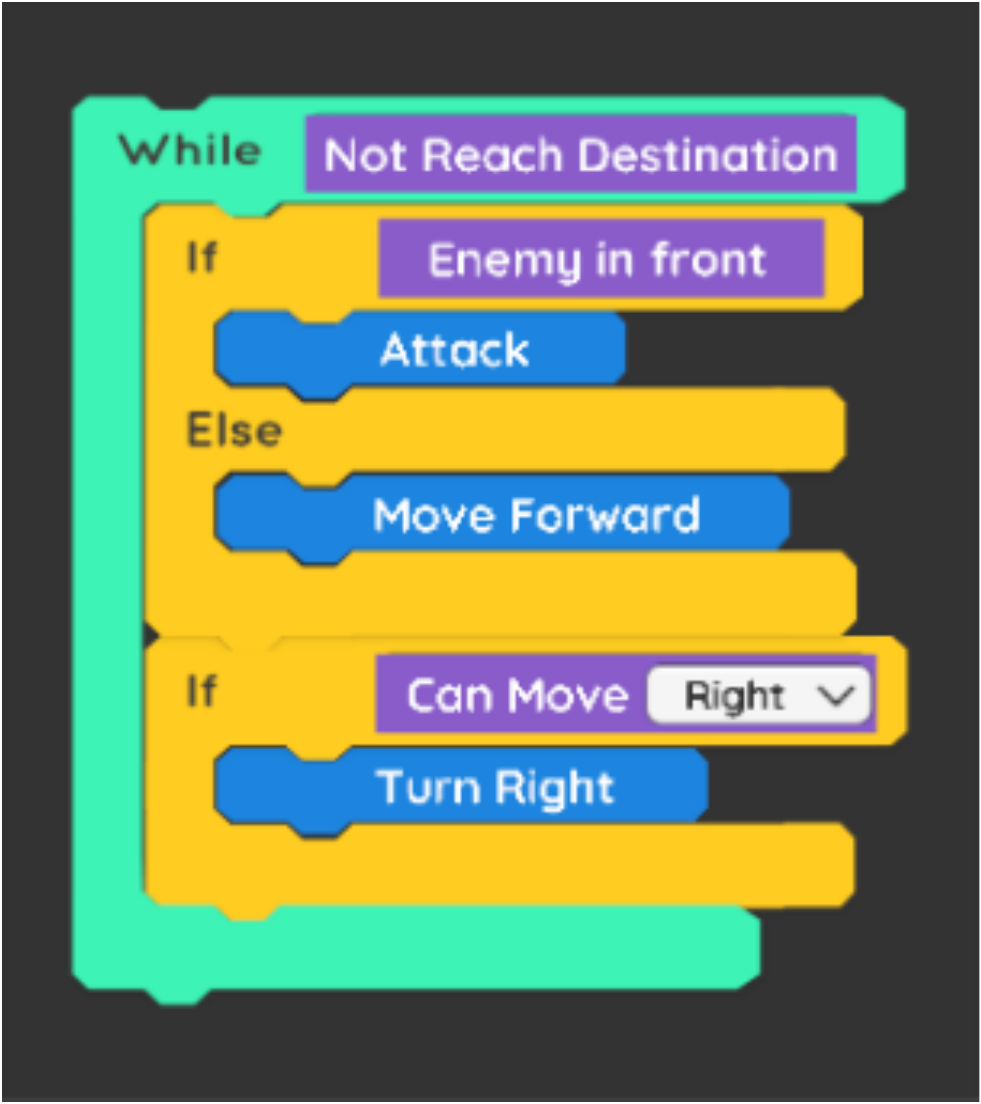}} \\
    \hline
  \end{tabular}
  \caption{Examples for low-efficiency and high-efficiency solutions of maze solving process in 3D maze system.}
  \label{fig:3dprint}
\end{table*}

\subsection{Prior Implementation of 3D Maze with Generative AI} \label{Section 3.2}
As mentioned above, the 3D Maze platform was designed to support students’ programming learning. The following section introduces a prior round of implementation at an Asian university. During this round of implementation, students were allowed to use all kinds of Generative AI to support the maze solving and maze design activities, as some students had limited programming experience. 

\subsubsection{Implementation Procedures and Data Collection}
The research team implemented the 3D Maze platform among 1500 undergraduate students at an Asian university in 2023. The course was an interdisciplinary course on digital literacy that was open to all first-year and second-year students from diverse backgrounds. The instructors use the 3D Maze platform to teach computational thinking and the emerging Generative AI in education. During the two-hour course session, students were required to complete the two activities described above: 1) design a maze containing different environmental assets and then 2) solve the maze using high-efficiency solutions. Considering that most students were first-year students with limited programming experience, they were allowed to use LLM tools freely to aid maze designs and maze solving activities. After the 3D maze activities, the students submitted reflection reports with the following questions on how Generative AI supports maze design and maze solving, respectively:
\begin{enumerate}
    \item Which Generative AI tools have you experimented with? Which one(s) did you find helpful? Why?
    \item For what kinds of subtasks (or purposes) have you used Generative AI? 
    \item In your opinion, what kinds of subtasks is the Generative AI competent in, and what is it not? 
    \item What prompt engineering techniques have you applied to elicit more helpful responses? Elaborate on your answers.
    \item In general, do you think Generative AI is helpful in maze design activity? Why?
\end{enumerate}

\subsubsection{Identified challenges from the previous round of implementation of the 3D Maze}
We conducted a content analysis \cite{elo2008qualitative} of students ' responses to understand how Generative AI was used to support their maze design and maze solving activities, and what the challenges. We identified three challenges that students reported in their reflection reports, which serve as the foundation for designing MazeMate.

{\textbf{Generative AI is not context-aware.} A recurring challenge students identified is that generative AI lacks awareness of the specific 3D maze context. While the tool can generate generic suggestions, it often requires students to spend extra effort carefully explaining the maze setup and rules in their prompts. As one student noted, ``\textit{we need to provide additional context or information in the input prompt (e.g., the maze was created and the legends). Even with this additional input, the AI sometimes produced suggestions that led to unsolvable mazes}”. 

\textbf{Inaccuracy of Generative AI outputs}. Students also emphasized the issue of inaccurate or unreliable outputs when solving complex maze problems. Although they provided detailed instructions, the AI frequently failed to generate correct solutions, with one student explaining that ``\textit{Generative AI is not competent to solve complicated maze problems. Given all the instructions, it still provides us with the wrong solution}.”

\textbf{Generative AI as cognitive shortcuts.} Some students have the tendency to use Generative AI as a cognitive shortcut. They valued the tool’s speed in producing new mazes or optimizing layouts, as one explained: \textit{``It can generate a new solvable maze very quickly”} and \textit{``Gen AI can optimize maze layouts for specific criteria, and can also create wide variety of maze designs.”} Others appreciated its creativity, saying, \textit{``Yes, Gen AI can be helpful in maze design by adding more creativity and challenges which we may not have thought of.”} 

Student feedback on using generative AI in 3D Maze activities highlighted both its benefits and drawbacks. Some students appreciated AI's speed and creativity, while others reported inaccurate results, poor contextual understanding, and a tendency to use AI as a shortcut instead of engaging in computational thinking. These issues highlighted the risk that Generative AI could hinder learning if left unstructured, echoing concerns reported by recent reviews of using LLM in programming learning \cite{pirzado2024navigating,raihan2025large}. These identified challenges have led to the development of MazeMate, an LLM-powered chatbot integrated into the 3D Maze environment. It is specifically designed to support maze solving and design, emphasizing the practice of computational thinking. 

\begin{figure*}
    \centering
    \includegraphics[width=1\linewidth]{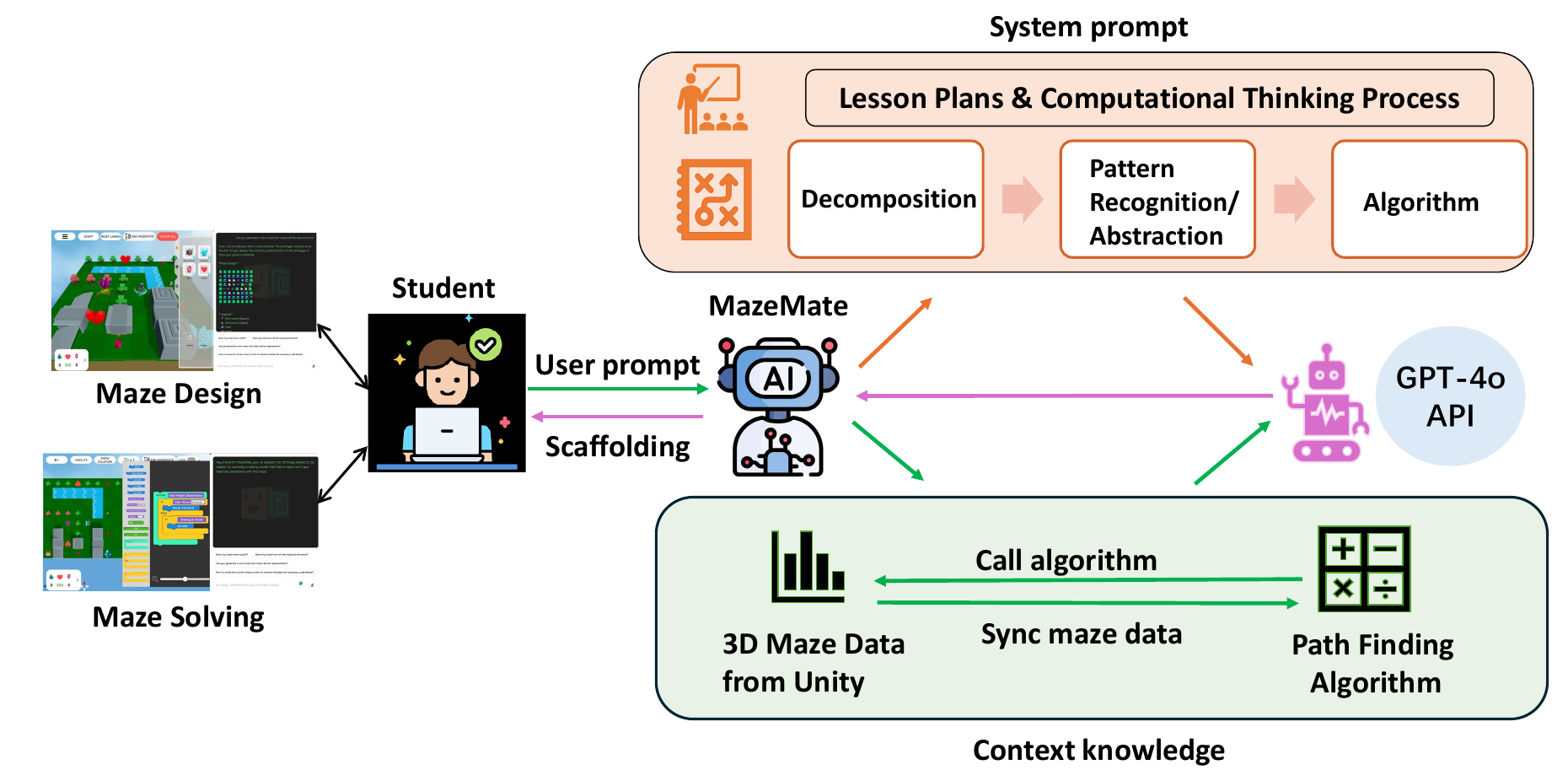}
    \caption{The overview of the interaction process between students and MazeMate.}
    \label{fig:overview}
\end{figure*}

\section{MazeMate Designs}

MazeMate is an enhanced LLM-driven tool in 3D maze system to foster students’ computational thinking skills in maze design and maze solving activities. Here, we utilize the GPT-4O API of OpenAI to support this system design. Fig.~\ref{fig:overview} shows the overview of the interaction process between students and MazeMate.
Drawing on findings from prior implementation of the 3D Maze, we identified four design requirements for MazeMate to operate as an effective pedagogical scaffold. First, MazeMate must demonstrate contextual awareness of the 3D Maze environment for adaptive supports, including constraints such as available design elements and pseudo-code rules, in order to provide relevant feedback. MazeMate should also synchronize game data to provide adaptive support in response to students’ progress. Second, the system should be supplied with correct solutions as contextual knowledge to generate tailored feedback, which requires additional path-finding algorithms as LLM in general cannot find a valid path for complex maze. Third, MazeMate should function as a pedagogical agent rather than a cognitive shortcut; to achieve this, prompting strategies were implemented to ensure that guidance is provided progressively. At the same time, effective scaffolding requires a deliberate alignment between MazeMate’s design features and the core processes of computational thinking. Thus, we introduce the design structures that explicitly integrate these design features with decomposition, pattern recognition, abstraction, and algorithmic thinking across maze solving and maze design, thereby ensuring their applicability in authentic classroom contexts.


\subsection{Technical Details of MazeMate} \label{Section 4}

\subsubsection{MazeMate Provides Contextualized and Adaptive Support}

MazeMate is tightly integrated with the 3D maze system, enabling continuous synchronization of live maze data to deliver context-aware instructional support. Before generating each response, MazeMate reads the student’s current maze state and tailors its feedback based on this real-time information.
Specifically, during the maze design activity, real-time data on the 3D Maze is synchronized with the MazeMate so that the MazeMate has contextual understanding. Student actions, such as placing obstacles, setting monsters, or modifying paths, are instantly transmitted to MazeMate, allowing it to provide timely suggestions throughout the design process (Fig.~\ref{fig:LLMMazeDesign}).
During the maze solving activity, as students attempt to solve the maze, MazeMate automatically captures their solution scripts and offers targeted advice based on their latest code (Fig.~\ref{fig:LLMMazesolving}).

\subsubsection{MazeMate Provides Accurate Solutions through Path Finding Algorithm} \label{Section 4.3}
Maze solving is a complex and logical task, relying on LLM alone to solve complex mazes is insufficient. The 3D Maze environment introduces layered game constraints (e.g., obstacle layouts, monster encounters, gem placements, and health resources), along with predefined pseudocode functions that MazeMate must adhere to. LLMs often fail to consistently account for these rules, leading to invalid or unsolvable paths.
To address this challenge, we integrate the LLM with two algorithmic modules to ensure that the MazeMate can provide accurate solutions within the 3D Maze environment: the BFS-based low-efficiency and the program-compression-based high-efficiency solutions. The role of path planning algorithms is twofold: on the one hand, it ensures that the designed maze is indeed solvable under the given constraints; on the other hand, it provides the logical foundation for generating stepwise and abstract solutions that scaffold students’ computational thinking process. Within MazeMate, the path planning algorithm is called when students explicitly request a solution to the maze, enabling the model to return either a low-efficiency or a high-efficiency solution depending on the stage of the computational thinking process (please see supplementary materials for the complete prompt). 

\textbf{Low-efficiency solution.} The low-efficiency algorithm is implemented using a breadth-first search (BFS). This method systematically explores the state space, including agent position, orientation, collected gems, and defeated monsters, to identify an optimal path that satisfies all constraints. The BFS returns a sequence of atomic actions such as Move Forward, Turn Left, or Attack. Presented as a step-by-step solution, this form of output helps students grasp the overall structure of the maze and serves as the basis for pattern recognition in the computational thinking process.

\textbf{High-efficiency solution.} The high-efficiency algorithm extends the basic BFS in a three-stage process. In the first stage, a feasible path that satisfies all constraints is obtained using BFS. In the second stage, the path sequence is transformed into a program tree structure by mapping consecutive actions and special events (e.g., encountering monsters or gems) to corresponding program nodes. Control structures such as conditional statements and loops are introduced when necessary, ensuring that the representation is semantically equivalent to the original path. In the third stage, a compression mechanism based on pattern recognition is applied to identify repeated structures within the program tree, which are then abstracted into loops or conditional constructs. If semantic deviations occur during compression, a patching procedure restores correctness. Additionally, a variable neighborhood search (VNS) strategy \cite{mladenovic1997variable} is incorporated to further refine the solution, balancing correctness, execution efficiency, and structural simplicity.

By leveraging these algorithmic backends, the system guarantees that every solution is feasible, semantically valid, and optimized without breaking the constraints of the 3D maze environment.
This hybrid design allows the LLM to delegate computationally intensive reasoning to backend algorithms, and thus focus on scaffolding dialogues for computational thinking. 

\begin{figure*}
    \centering
    \includegraphics[width=1\linewidth]{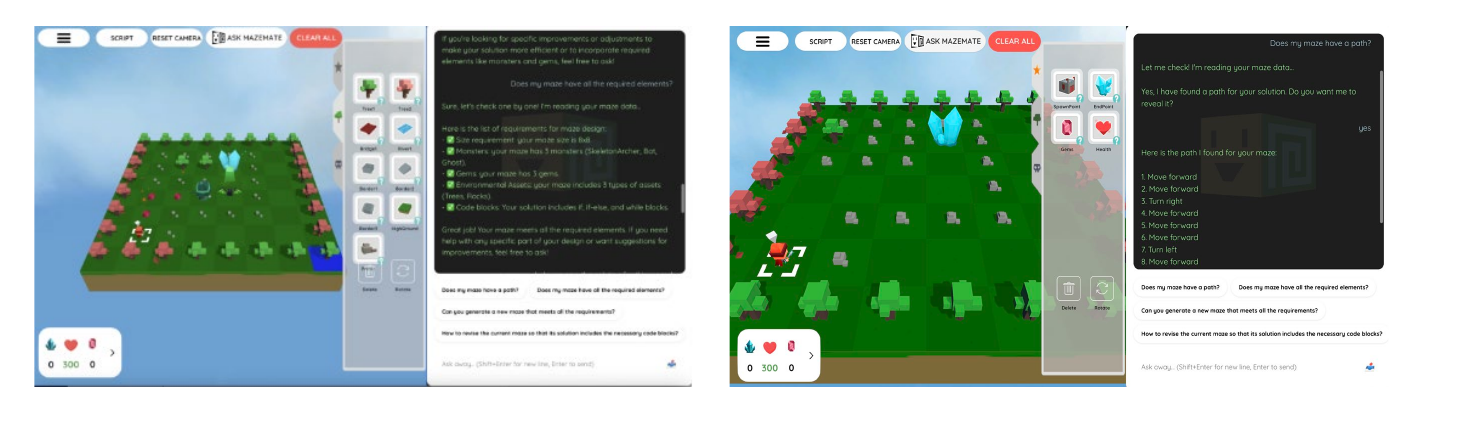}
    \caption{Maze design process with the help of MazeMate in 3D maze system.}
    \label{fig:LLMMazeDesign}
\end{figure*}

\begin{figure*}
    \centering
    \includegraphics[width=1\linewidth]{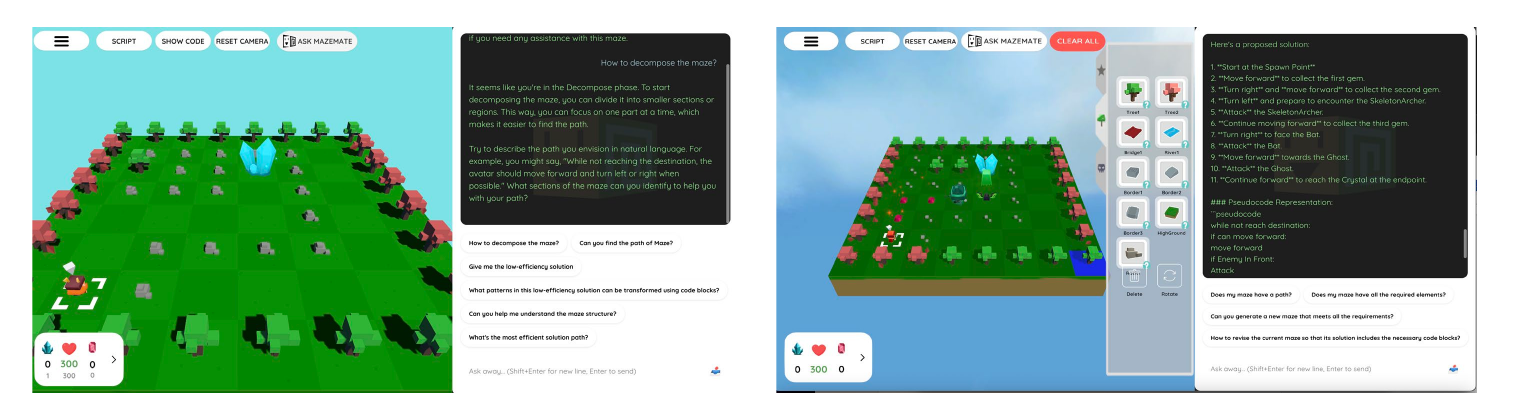}
    \caption{Maze solving process with the help of MazeMate in 3D maze system.}
    \label{fig:LLMMazesolving}
\end{figure*}

\subsubsection{Prompting for Progressive Scaffolding of MazeMate}
To prevent MazeMate from becoming a cognitive shortcut, the system is designed to guide students through the maze solving process gradually rather than providing immediate solutions. We formulated the requirement of this staged reveal of answers as prompts to feed into LLM (please see supplementary materials for the complete prompt). When students ask for the solutions to the maze, MazeMate will use a staged disclosure mechanism that unfolds across three interactions. On the first request for an answer, MazeMate generates a low-efficiency solution as step-by-step actions, and prompts students to grasp the overall path structure and practice problem decomposition in natural language. On the second request, MazeMate provides hints that introduce transformation rules that encourage students to convert stepwise paths into more efficient logical statements using condition blocks such as while, if, or repeat, thereby bridging natural reasoning with algorithmic representation. Only on the third request does MazeMate provide a high-efficiency solution expressed in algorithmic form. This design aims to reinforce computational thinking while keeping the AI as an effective scaffolding tool. 

\subsection{Alignment of MazeMate Scaffolds with Computational Thinking Processes}
Given that 3D Maze is designed for authentic classrooms to help students practice computational thinking, the scaffoldings provided by MazeMate need to be pedagogically aligned with the lesson plans. We obtained the lesson plan from the instructors and then further tailored the MazeMate scaffolds based on the lesson plan (Please see the supplementary materials for the lesson plans).  In the following section, we will describe how MazeMate’s features were integrated with maze solving and maze design activities to support computational thinking. MazeMate’s responses are capped at ten sentences per turn and framed in a scaffolding-oriented style that emphasizes questioning and explanation.

\subsubsection{Maze Solving}
As presented in Figure~\ref{fig:CT_solving}, the classroom activity introduces 3D Maze features through three predetermined mazes, each highlighting specific mechanics. For example, Maze B requires players to collect gems before reaching the goal and introduces conditional blocks. Across these tasks, students are instructed to aim for high-efficiency solutions, and MazeMate scaffolds their progress by aligning gameplay with computational thinking (CT) processes.
In the \textbf{Decomposition stage}, MazeMate encourages students to describe the path in natural language (e.g., \textit{“move forward three steps, then turn right”}). This helps them break the maze into manageable steps and recognize that complex navigation can be reduced to simpler actions. In the \textbf{Pattern Recognition \& Abstraction stage}, the LLM helps students notice repetition and conditional structures within the low-efficiency solution. For instance, repeated sequences of “move forward” can be abstracted into a while loop, and decision points can be mapped to if-statements. By guiding students to express these patterns in algorithmic terms, the model fosters abstraction skills and promotes the transition from specific instances to generalized solutions. In the\textbf{ Algorithm stage}, the LLM prompts students to synthesize their insights into a high-efficiency solution expressed as structured pseudocode.  

\begin{figure*}
    \centering
    \includegraphics[width=1\linewidth]{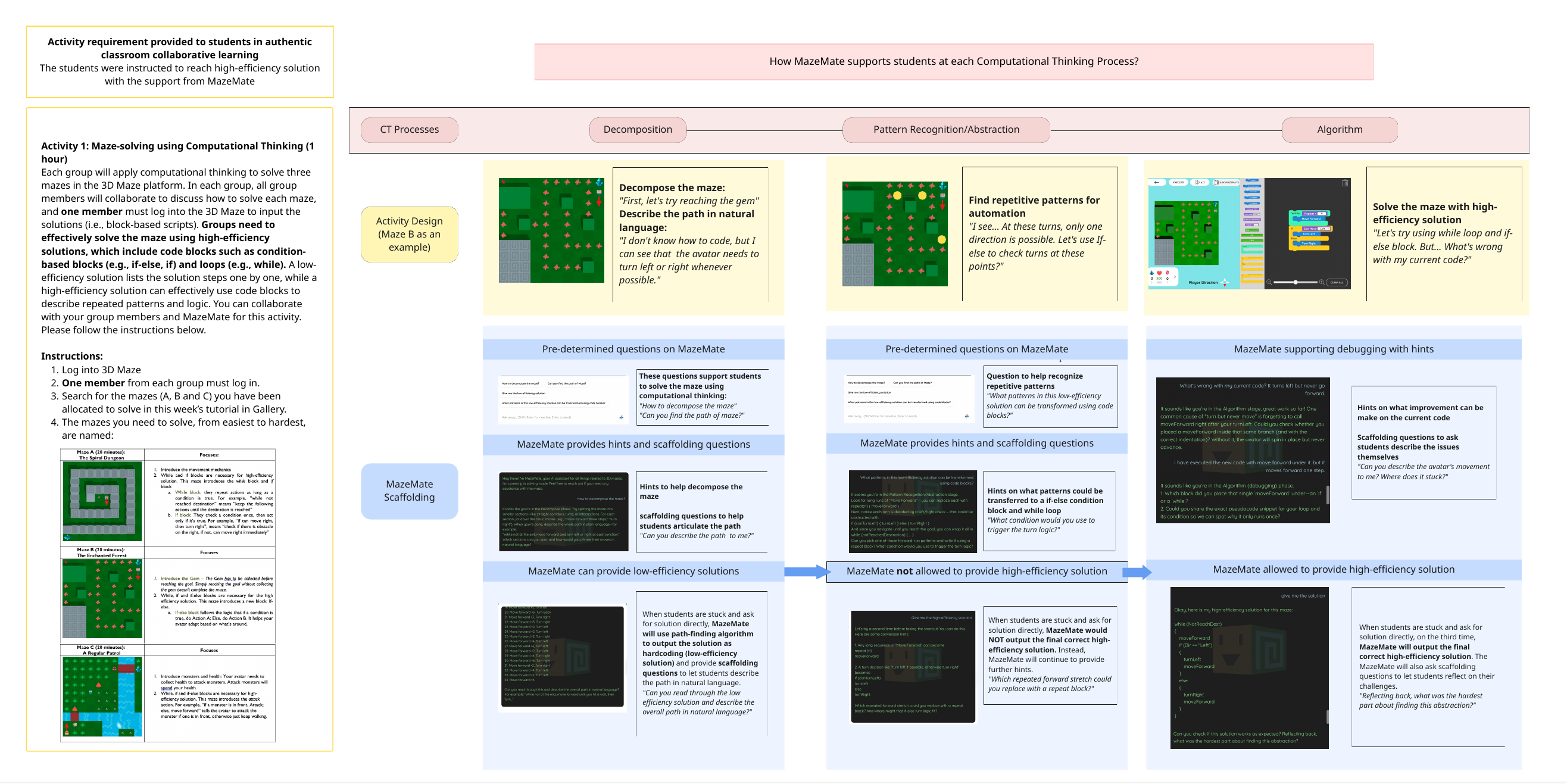}
    \caption{The design structure of aligning MazeMate with computational thinking in maze solving.}
    \label{fig:CT_solving}
\end{figure*}

\subsubsection{Maze Design}
In the maze design activity, the MazeMate is configured to scaffold students’ efforts in refining their creations while adhering to predefined design requirements, as presented in Figure\ref{fig:CT_Design}. The activity requirement asks students to be creative by designing the maze with a storyline (e.g., a dragon trap guarding the treasure) and including diverse environmental assets. To foster computational thinking, the lesson plan requires students to consider the algorithm blocks needed for their design.

In the \textbf{Decompose stage,} MazeMate checks whether the maze satisfies basic criteria (e.g., number of monsters and gems, diversity of assets) and encourages students to break their ideas into smaller design elements such as asset types and monster-health allocation. This helps students analyze the maze systematically and understand how each element contributes to solvability and playability. MazeMate can also call the path-finding algorithm (as explained in Section \ref{Section 4.3}) to verify whether the designed maze contains a valid solution path. In the \textbf{Pattern Recognition \& Abstraction stage}, the LLM guides students to identify repeated structural features, such as junctions, dead ends, or U-turn, and to connect these structures with corresponding algorithmic expressions (e.g., while-loops for repetitive paths, if-statements for conditional turns). Through this process, students are encouraged to represent complex maze patterns with algorithm blocks, preparing them to find the high-efficiency solution of the designed maze. In the \textbf{Algorithm stage}, the LLM helps students to find a high-efficiency algorithm for the designed maze. Rather than offering explicit solutions, it prompts students to consider how algorithmic blocks (while, if, if-else) would be applied, and reveals the high-efficiency solution upon students' request by calling the path-finding algorithm.

\begin{figure*}
    \centering
    \includegraphics[width=1\linewidth]{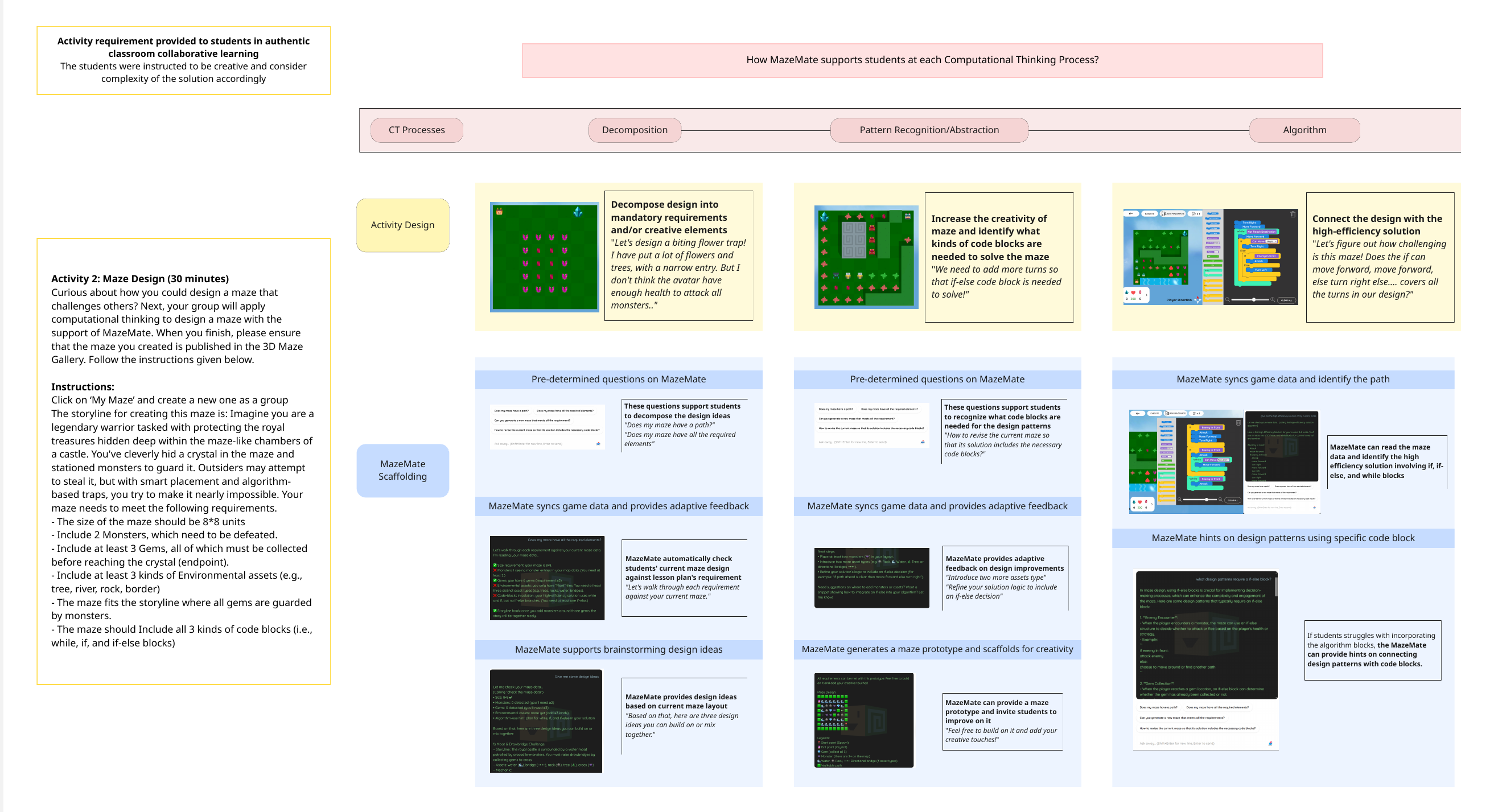}
    \caption{The design structure of aligning MazeMate with computational thinking in maze design.}
    \label{fig:CT_Design}
\end{figure*}

\section{User Study} \label{Section 5}
We conducted a user study in an authentic classroom to evaluate the effectiveness of MazeMate in fostering computational thinking across maze solving and maze design activities. To guide our investigation, we have the following research questions (RQs): 
\begin{enumerate}
    \item How do students perceive the helpfulness of MazeMate in maze design and maze solving, respectively?
    \item How do students perceive MazeMate supporting their computational thinking in maze design and maze solving, respectively?
    \item How does the perceived helpfulness of MazeMate relate to learning outcomes in maze design and maze solving, respectively? 
\end{enumerate}

\subsection{Participants and Context}
The participants of this study were first- and second-year students (enrolled in the university in 2023 and 2024) in a digital literacy course in the August semester of 2025 at an Asian university. In total, around 250 students across various schools in the university enrolled in this digital literacy course, and they were divided into tutorials, with around 20 students in each tutorial. In each tutorial, 4 to 5 students from different schools were purposefully assigned to a group that would work on various problem-solving activities in class. The project received approval from the university’s ethics review board to waive consent forms from the participants because the course took place in a regular educational setting, and the risk was minimal. Eventually, we collected data from 247 students in the user study. Among these students, 165 indicated that they used MazeMate during the solving process, and 166 indicated that they used MazeMate during the design process.

\subsection{Study Procedures}
The study took around one and a half hours and took place in an authentic classroom. It involved two structured activities integrating computational thinking. First, in the maze solving task, members in the group collaboratively solved three progressively complex mazes on the 3D Maze platform, employing efficient block-based programming (e.g., while, if, if-else) to solve the maze (Please see fig. \ref{fig:CT_solving}). Second, in the maze design task, groups created and published an original 8×8 maze, embedding narrative elements, monsters, gems, and environmental assets, while ensuring the maze required algorithmic reasoning and the application of key programming constructs (Please see fig. \ref{fig:CT_Design}).  

\subsection{Data Collection and Analysis}
After the participants engage in two 3D Maze activities, they fill in a reflection survey about their learning experiences engaging with MazeMate. The reflection survey includes students’ perceived helpfulness of MazeMate, their open-ended reflection on how MazeMate supports computational thinking, and the quiz on measuring learning outcomes. We collected 247 survey responses. 

For \textbf{the perceived helpfulness of MazeMate}, each participant answered two questions on a Likert Scale:  “How helpful is the trained GPT, MazeMate, in maze solving, on a scale of 1 (not helpful at all) to 10 (very helpful)?” and “How helpful is the trained GPT, MazeMate, in maze design, on a scale of 1 (not helpful at all) to 10 (very helpful)?”. To understand the relationship between perceived helpfulness and \textbf{learning outcome}, we further measure students’ learning through 8 quiz items. This formative quiz includes conceptual understanding of computation thinking such as ``What computational thinking concept help you identify repeated code so to simplify the solution using code blocks?", code logic such as ``What type of logic is this: `if there is a monster in front, attack; otherwise, keep moving'?", and transitions between hardcoding and high-efficiency code with code logic such as ``What is the high-efficiency solution for this `Move Forward*3, Turn Right, Turn Back, Move Forward*6, Turn Right'?". To understand \textbf{how MazeMate supports computational thinking}, we ask two open-ended questions: ``How did the trained GPT, MazeMate, support your computational thinking in maze design?” and “How did the trained GPT, MazeMate, support your computational thinking in maze solving?”. (Please see supplementary materials for complete quiz items and survey questions). 

To answer the first research question, we conducted descriptive analyses on their perceived helpfulness for maze design and maze solving activities, respectively. Second, for the research question about how the perceived helpfulness correlates with learning outcome, we conducted a correlation analysis between quiz scores and perceived helpfulness for maze solving and maze design, respectively. Lastly, we conducted a thematic analysis of students’ answers to two open-ended questions on how MazeMate supported computational thinking in maze solving and maze design. Following the procedure of thematic analysis, we first mapped students’ responses to the core processes of computational thinking, and then grouped similar responses into themes within each process \cite{vaismoradi2013content}. Students also reported problems when interacting with MazeMate, which were categorized separately.

\section{Findings}
The distribution of perceived helpfulness of MazeMate for maze design and maze solving is presented in Figure.\ref{fig:bar}. On average, students rated MazeMate as moderately helpful for both maze solving (Mean$=6.15$, Standard Deviation$=2.36$) and maze design (M$=5.96$, SD$=2.55$), on a 10-point scale. Perceived helpfulness was slightly higher for maze solving than for maze design. Ratings clustered around the midpoint in both cases, with some students reporting that MazeMate was highly supportive and others finding its contributions limited. These findings suggest that while MazeMate was generally seen as useful, students experienced its benefits unevenly across different activities.

\begin{figure*}[ht]
    \centering
    \includegraphics[width=0.75\linewidth]{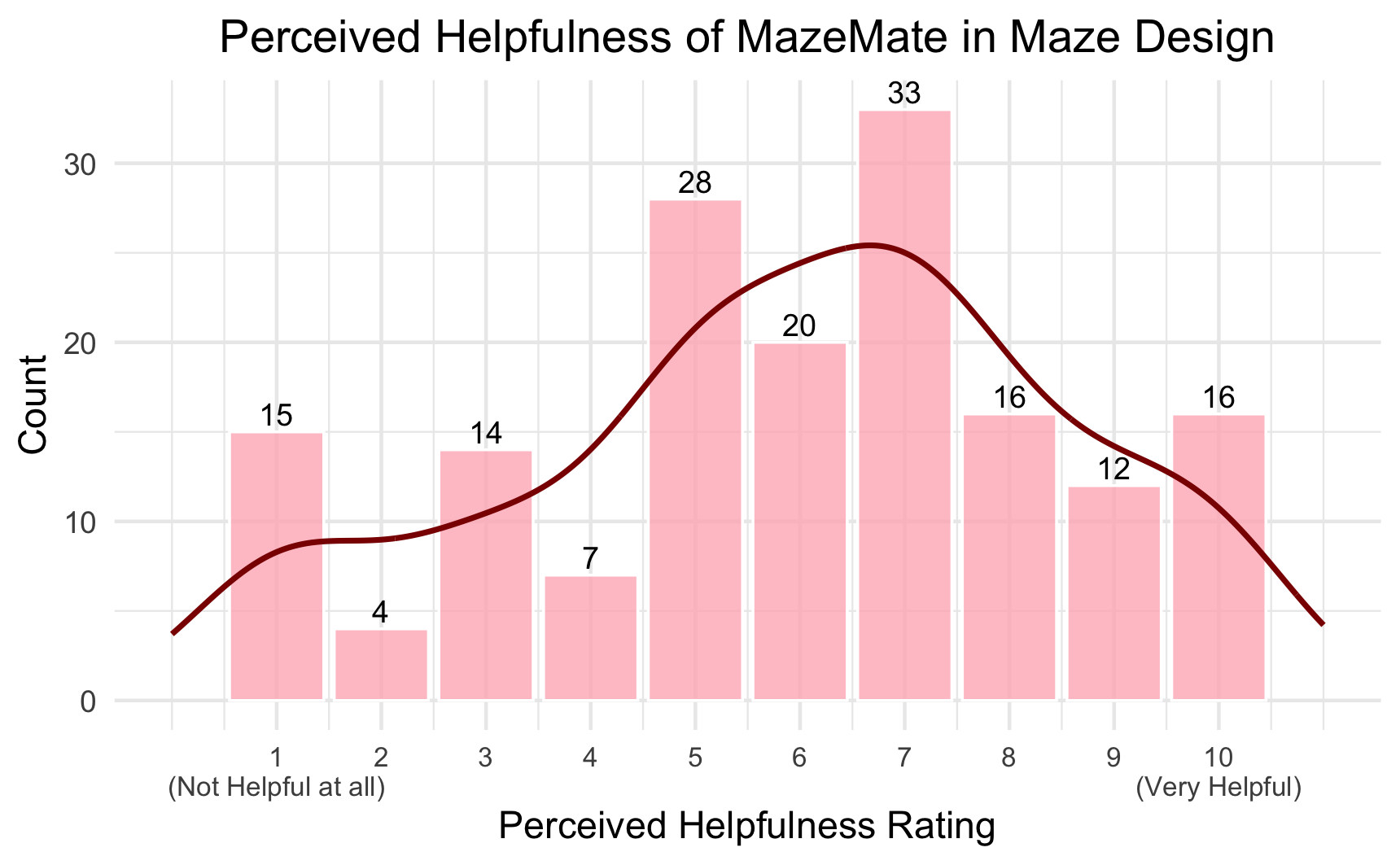}\\[1ex]
    \includegraphics[width=0.75\linewidth]{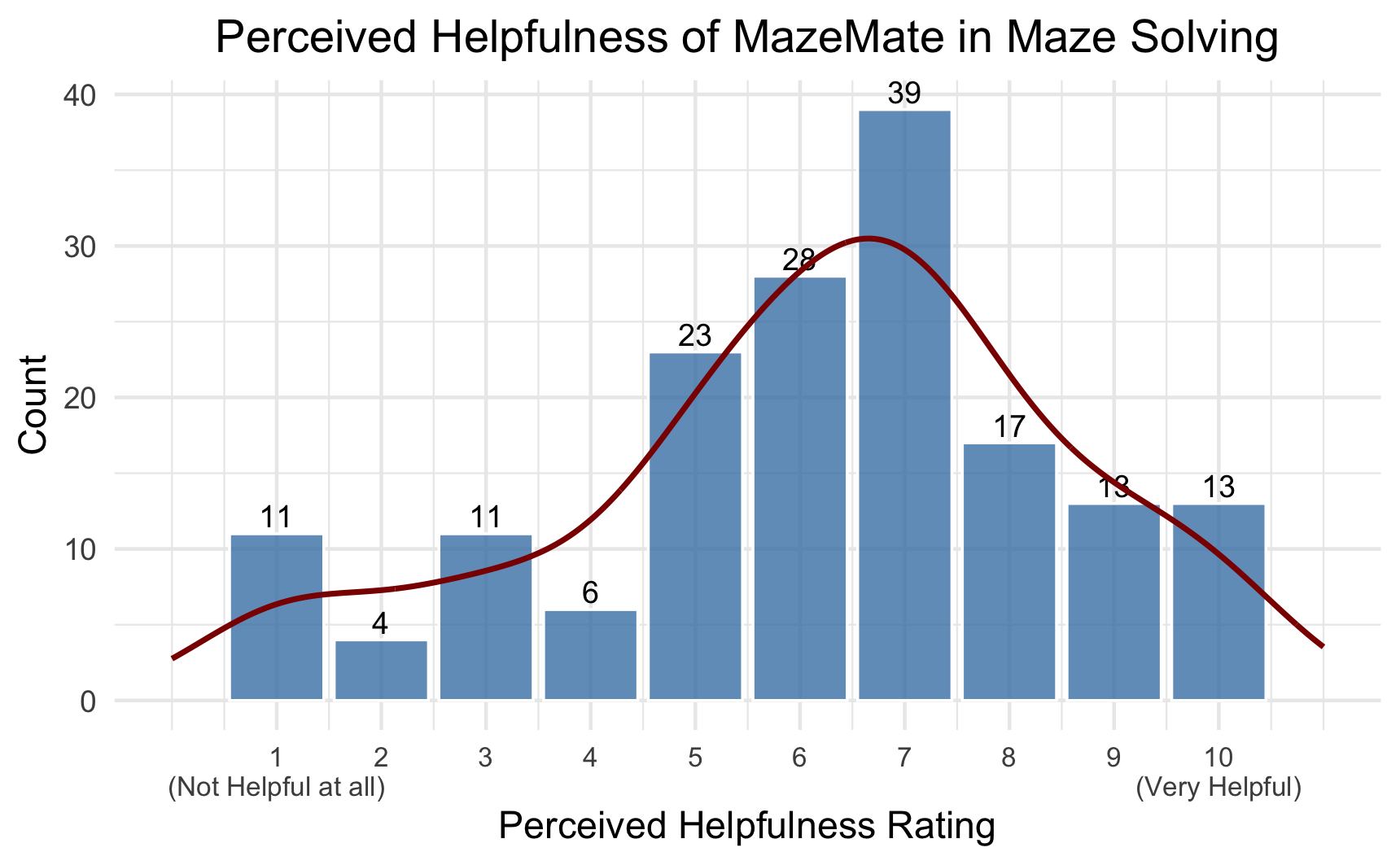}
    \caption{The distribution of perceived helpfulness of MazeMate for maze design and maze solving.}
    \label{fig:bar}
\end{figure*}

A small positive correlation was observed between perceived helpfulness of MazeMate in maze solving and quiz scores ($\rho = 0.15, p = 0.054$), suggesting a trend where students who viewed MazeMate as more helpful tended to achieve higher scores, though this relationship is marginally significant. In contrast, perceived helpfulness in maze design showed no meaningful association with quiz performance ($\rho= 0.08, p= 0.29$). These findings indicate that MazeMate’s perceived support in solving tasks may be more closely tied to measurable learning outcomes than its role in design activities.

Lastly, thematic analysis of student responses revealed how MazeMate supported CT across maze solving and maze design activities. In maze design, one theme related to decomposition was \textbf{checking maze requirements}. Students noted that MazeMate helped them verify whether their designs satisfied the given constraints. As one student explained, “It told us if our maze met the requirements for design,” while another added, “Sorting out to check if criteria are met.” For pattern recognition and abstraction, the theme \textbf{feedback to generate alternative design ideas} captured how MazeMate provided suggestions that inspired new possibilities and encouraged students to view their mazes from multiple perspectives. Students shared, “It gave me ideas on how to create a maze,” and “Helps me think from different perspectives so that my maze has different varieties and challenges.” Providing feedback to refine designs illustrated how MazeMate strengthened students’ design logic, for example, “It gave us valuable feedback contributing to the improvement of the maze design.” and “My group was stuck with a situation where the maze was unsolvable, and the GPT provided hints to fix it.”

In maze solving, decomposition was supported through \textbf{decomposing solutions into smaller steps}, as students described how MazeMate broke problems into manageable parts: \textit{“It broke down the maze into small problems which made it easier to solve”} and \textit{“It helped me break down the maze solving by providing step-by-step instructions.”} Another theme, \textbf{exploring the efficiency of solutions}, reflected how students used MazeMate to compare alternative approaches: \textit{“MazeMate helped in creating low efficiency solutions and refining them.”} For pattern recognition and abstraction, two themes were identified. \textbf{Identifying and simplifying repeated patterns} described how MazeMate encouraged students to eliminate redundancies, e.g., \textit{“It helps me to eliminate repeated steps into more efficient solutions.”} \textbf{Offering hints towards a high-efficiency solution} captured its ability to suggest new strategies, such as loops and conditionals: \textit{“MazeMate could give information on how to use loops and conditions for efficiency.”} In algorithmic thinking, students described two areas of support: \textbf{providing correct algorithms}, as one noted, \textit{“It provided us with the algorithm that could be used to solve the maze,”} and \textbf{debugging solutions}, as another remarked, \textit{“It told me what could be the issue for code that didn’t work.”}

Despite these contributions, students also identified challenges. The first theme, \textbf{mismatched suggestions}, referred to cases where MazeMate generated responses that did not align with the questions or failed to address the intended task. For example, one student shared, \textit{“We tried to ask it to make an impossible maze, but it didn’t respond effectively,”}. This issue often arose when students deliberately posed questions outside the task requirements, such as requesting an impossible maze. In such cases, MazeMate returned irrelevant information or generic statements like \textit{“AI not configured,”} which compromised students’ experiences. Another noted, \textit{“Not much, it would be better if it can find problems in our solutions directly.”} This limitation stemmed from MazeMate’s design as a scaffold: it was configured to provide hints for improving a solution (e.g., moving from low-efficiency to high-efficiency paths) rather than generating fully correct answers. While pedagogically intended, this sometimes conflicted with students’ expectations of directly pinpointing the errors and suggesting the fix.

The second theme, \textbf{fabricated algorithm solutions}, reflected frustrations with irrelevant or unusable outputs. Students reported, \textit{“[MazeMate] did not help, gave a solution that didn’t work,”} and \textit{“It gave us some answers but wasn’t very accurate.”} These glitches often occurred when MazeMate produced commands that did not exist in the 3D Maze environment. They appeared more frequently in maze design tasks, where extended interactions or increasingly complex, self-designed mazes exceeded MazeMate’s ability to generate reliable, high-efficiency solutions.

Overall, these findings show that MazeMate provided scaffolding aligned with CT processes by supporting decomposition, abstraction, and algorithmic thinking, while also revealing important limitations. Based on students' perceived helpfulness, its correlation with learning outcomes, and the thematic analysis, MazeMate is more useful for maze solving than for maze design. These analyses highlight the need to improve AI scaffolds to balance accuracy, contextual relevance, and student engagement in CT development.

\section{Discussion}
This study presented the design and evaluation of MazeMate, an LLM-powered chatbot embedded in the 3D Maze programming environment to scaffold computational thinking (CT). MazeMate was designed with three design features: contextual awareness of the 3D Maze environment and adaptive scaffolding responsive to student progress, integration of correct solutions as contextual knowledge, and staged guidance to avoid cognitive shortcuts. These features were then integrated with lesson plans and computational thinking processes to provide scaffolding for maze solving and maze design tasks in the 3D Maze environment. Findings from the user study demonstrate that MazeMate can provide meaningful scaffolding aligned with CT processes, particularly decomposition, pattern recognition/abstraction, and algorithmic thinking in maze solving.
The thematic analysis revealed that, in maze solving task, MazeMate helped them break down the complex maze into smaller steps, identify and simplify repeated patterns, and debug or refine algorithms. In maze design task, MazeMate also supported abstraction by suggesting alternative design ideas and connecting designs with code blocks, though students highlighted more limitations in this activity. Together, these findings indicate that MazeMate has potential as a pedagogical agent to support students in practicing the fundamental dimensions of CT.

Perceived helpfulness ratings suggest that MazeMate was regarded as slightly more supportive in maze solving than in maze design. Importantly, perceived helpfulness in maze solving showed a marginally significant positive correlation with learning outcome, suggesting that MazeMate’s scaffolding in maze solving might have contributed to learning gains. By contrast, no such relationship was observed for maze design, and student reflections reported more problems with MazeMate, such as mismatched suggestions in maze design. A likely explanation is that MazeMate, built on general-purpose LLMs with domain-specific knowledge provided only through prompts, struggles to maintain contextual accuracy in creative design tasks \cite{nesbitt2024semantic,welz2024enhancing,leinonen2023using}. Another factor is the balance between delivering scaffolding and providing solutions. While staged guidance was designed to prevent students from using MazeMate as cognitive shortcuts, it might have conflicted with some students' expectations that the AI responses should always be direct and address the question. Thus, when MazeMate did not provide immediately usable answers, they abandoned AI assistance. This tension underscores the challenge of designing AI scaffolds that encourage sustained engagement and practice of computational thinking. 

Based on these insights, several design implications emerge. We organize these into areas where MazeMate has achieved and areas where further refinement is needed.

Our designed MazeMate performs well on logic-intensive, highly structured tasks with clear quantitative evaluation criteria, such as maze solving in the 3D maze environment. We implement it through path planning algorithms at different stages in the CT process, as referred to in the techniques details of MazeMate. It can reduce hallucinations and appropriately guide the full CT process. Specifically, First, MazeMate demonstrates a clear linkage between scaffolds and computational thinking, functioning as a pedagogical scaffold that helps students decompose tasks, recognize patterns, and debug solutions. Second, MazeMate can generate contextualized and adaptive responses by drawing on real-time game data to provide more personalized guidance. Third, the staged reveal of answers ensures that the MazeMate can eventually provide accurate solutions, while still encouraging students to engage in CT processes before receiving the final outcome.

However, the results also show that in open-ended tasks that emphasize novelty and creativity, such as 3D maze design, MazeMate, and more generally current generative AI approaches \cite{jiang2024survey,bellemare2024divergent}, continue to exhibit hallucination issues. A key reason may be the absence of quantitative guidelines to evaluate and steer the accuracy of LLM-generated outputs in subjective, creativity-oriented tasks. Given that this was MazeMate’s first-round evaluation, our next step is to focus on mitigating AI hallucinations in open-ended tasks that emphasize novelty and creativity.  

Further, an area of improvement is strengthening scaffolds to elicit reasoning from students. For example, MazeMate can require students to articulate their logic before receiving advanced hints. MazeMate also needs to provide better support for creative design tasks, guiding narrative construction or design complexity without overshadowing students’ own creativity. Another area of improvement is to manage student impatience with staged answers by adding mechanisms such as intermediate feedback (e.g., "before I tell you the answer, let's try if you can solve the maze with the following hints"), or visible progress indicators to sustain motivation while maintaining the pedagogical intent of progressive scaffolding.

\section{Limitations}
This study has several limitations that should be acknowledged. First, the evaluation of MazeMate was conducted within a single two-hour classroom session. While this setting provided an authentic glimpse into how students interacted with the system, it may not have allowed students sufficient time to practice the full process of computational thinking. Future research should involve longitudinal studies across multiple sessions to capture how sustained use of MazeMate supports skill development.

Second, the evaluation relied primarily on students’ self-reported perceptions of MazeMate’s helpfulness. Although these reflections offer valuable insights into students’ experiences, they may be subject to bias. To address this limitation, future work should incorporate more objective measures such as interaction log data, code submissions, or analysis of problem-solving processes to provide richer evidence of how MazeMate supports computational thinking.

Third, the user study did not include a traditional experimental baseline for comparison. Creating this baseline in real classroom teaching would have involved not providing scaffolding to certain students, which was considered ethically problematic. Nevertheless, future research could explore alternative experimental designs, such as rotating access to specific features or using historical control data, to evaluate MazeMate’s impact more rigorously.

\section{Conclusion}
This paper introduced MazeMate, an LLM-powered chatbot embedded in a gamified 3D Maze programming environment, designed to scaffold the development of computational thinking (CT). Unlike general-purpose large language models that often provide direct answers without considering pedagogical objectives, MazeMate is designed with three design features. First, MazeMate has contextual awareness of the 3D Maze environment to provide adaptive scaffolding responsive to students’ progress. MazeMate synchronizes the game data before every response to students' queries to align its responses with the constraints of the 3D Maze system and students' progress. Second, we integrate path-finding algorithms into MazeMate to reduce LLM hallucinations in complex and logical tasks (i.e., maze solving in the 3D Maze). 
Lastly, MazeMate is designed to provide staged guidance to prevent it from being a cognitive shortcut. Furthermore, we align these design features with computational thinking and lesson plans to ensure that MazeMate could serve as a pedagogical agent, guiding students to actively engage with CT processes.

We presented the first round of implementation of MazeMate with 247 undergraduates in authentic classrooms. The results showed that MazeMate can scaffold key CT practices in meaningful ways. However, students also identified challenges, such as mismatched suggestions and fabricated algorithm solutions in maze design tasks that are open-ended and creative. Thus, our subsequent iterations of MazeMate will focus on enhancing contextual grounding of LLM in the 3D maze environment, improving transparency of AI responses to maintain sustained engagement in CT processes, and extending support for creative design tasks. These refinements will further ensure that an AI-supported gamified programming environment, such as a 3D maze, can foster the deeper computational thinking and problem-solving skills that are essential for students in the 21st century.

\section{Acknowledgements of the Use of AI}
We have utilized GPT-4o for prototyping and deploying the MazeMate, with details explained in Section 4. All the authors take responsibility for the use of Generative AI in this paper.

\bibliographystyle{ACM-Reference-Format}
\bibliography{sample-sigconf-authordraft,software}


\begin{thebibliography}{49}


\ifx \showCODEN    \undefined \def \showCODEN     #1{\unskip}     \fi
\ifx \showISBNx    \undefined \def \showISBNx     #1{\unskip}     \fi
\ifx \showISBNxiii \undefined \def \showISBNxiii  #1{\unskip}     \fi
\ifx \showISSN     \undefined \def \showISSN      #1{\unskip}     \fi
\ifx \showLCCN     \undefined \def \showLCCN      #1{\unskip}     \fi
\ifx \shownote     \undefined \def \shownote      #1{#1}          \fi
\ifx \showarticletitle \undefined \def \showarticletitle #1{#1}   \fi
\ifx \showURL      \undefined \def \showURL       {\relax}        \fi
\providecommand\bibfield[2]{#2}
\providecommand\bibinfo[2]{#2}
\providecommand\natexlab[1]{#1}
\providecommand\showeprint[2][]{arXiv:#2}

\bibitem[Angeli and Giannakos(2020)]%
        {angeli2020computational}
\bibfield{author}{\bibinfo{person}{Charoula Angeli} {and} \bibinfo{person}{Michail Giannakos}.} \bibinfo{year}{2020}\natexlab{}.
\newblock \bibinfo{title}{Computational thinking education: Issues and challenges}.
\newblock \bibinfo{numpages}{106185}~pages.
\newblock


\bibitem[Banic and Gamboa(2019)]%
        {banic2019visual}
\bibfield{author}{\bibinfo{person}{Amy Banic} {and} \bibinfo{person}{Ruben Gamboa}.} \bibinfo{year}{2019}\natexlab{}.
\newblock \showarticletitle{Visual design problem-based learning in a virtual environment improves computational thinking and programming knowledge}. In \bibinfo{booktitle}{\emph{2019 IEEE Conference on virtual reality and 3D User Interfaces (VR)}}. IEEE, \bibinfo{pages}{1588--1593}.
\newblock


\bibitem[Basawapatna et~al\mbox{.}(2013)]%
        {basawapatna2013simulation}
\bibfield{author}{\bibinfo{person}{Ashok~Ram Basawapatna}, \bibinfo{person}{Alexander Repenning}, {and} \bibinfo{person}{Clayton~H Lewis}.} \bibinfo{year}{2013}\natexlab{}.
\newblock \showarticletitle{The simulation creation toolkit: an initial exploration into making programming accessible while preserving computational thinking}. In \bibinfo{booktitle}{\emph{Proceeding of the 44th ACM technical symposium on Computer science education}}. \bibinfo{pages}{501--506}.
\newblock


\bibitem[Bellemare-Pepin et~al\mbox{.}(2024)]%
        {bellemare2024divergent}
\bibfield{author}{\bibinfo{person}{Antoine Bellemare-Pepin}, \bibinfo{person}{Fran{\c{c}}ois Lespinasse}, \bibinfo{person}{Philipp Th{\"o}lke}, \bibinfo{person}{Yann Harel}, \bibinfo{person}{Kory Mathewson}, \bibinfo{person}{Jay~A Olson}, \bibinfo{person}{Yoshua Bengio}, {and} \bibinfo{person}{Karim Jerbi}.} \bibinfo{year}{2024}\natexlab{}.
\newblock \showarticletitle{Divergent creativity in humans and large language models}.
\newblock \bibinfo{journal}{\emph{arXiv preprint arXiv:2405.13012}} (\bibinfo{year}{2024}).
\newblock


\bibitem[Brennan and Resnick(2012)]%
        {brennan2012new}
\bibfield{author}{\bibinfo{person}{Karen Brennan} {and} \bibinfo{person}{Mitchel Resnick}.} \bibinfo{year}{2012}\natexlab{}.
\newblock \showarticletitle{New frameworks for studying and assessing the development of computational thinking}. In \bibinfo{booktitle}{\emph{Proceedings of the 2012 annual meeting of the American educational research association, Vancouver, Canada}}, Vol.~\bibinfo{volume}{1}. \bibinfo{pages}{25}.
\newblock


\bibitem[Burke et~al\mbox{.}(2016)]%
        {burke2016computational}
\bibfield{author}{\bibinfo{person}{Quinn Burke}, \bibinfo{person}{W~Ian O'Byrne}, {and} \bibinfo{person}{Yasmin~B Kafai}.} \bibinfo{year}{2016}\natexlab{}.
\newblock \showarticletitle{Computational participation: Understanding coding as an extension of literacy instruction}.
\newblock \bibinfo{journal}{\emph{Journal of adolescent \& adult literacy}} \bibinfo{volume}{59}, \bibinfo{number}{4} (\bibinfo{year}{2016}), \bibinfo{pages}{371--375}.
\newblock


\bibitem[Cambaz and Zhang(2024)]%
        {cambaz2024use}
\bibfield{author}{\bibinfo{person}{Doga Cambaz} {and} \bibinfo{person}{Xiaoling Zhang}.} \bibinfo{year}{2024}\natexlab{}.
\newblock \showarticletitle{Use of ai-driven code generation models in teaching and learning programming: a systematic literature review}. In \bibinfo{booktitle}{\emph{Proceedings of the 55th ACM Technical Symposium on Computer Science Education V. 1}}. \bibinfo{pages}{172--178}.
\newblock


\bibitem[Chang(2014)]%
        {chang2014effects}
\bibfield{author}{\bibinfo{person}{Chih-Kai Chang}.} \bibinfo{year}{2014}\natexlab{}.
\newblock \showarticletitle{Effects of using Alice and Scratch in an introductory programming course for corrective instruction}.
\newblock \bibinfo{journal}{\emph{Journal of Educational Computing Research}} \bibinfo{volume}{51}, \bibinfo{number}{2} (\bibinfo{year}{2014}), \bibinfo{pages}{185--204}.
\newblock


\bibitem[Cheng et~al\mbox{.}(2017)]%
        {cheng2017teaching}
\bibfield{author}{\bibinfo{person}{Alan Cheng}, \bibinfo{person}{Lei Yang}, {and} \bibinfo{person}{Erik Andersen}.} \bibinfo{year}{2017}\natexlab{}.
\newblock \showarticletitle{Teaching language and culture with a virtual reality game}. In \bibinfo{booktitle}{\emph{Proceedings of the 2017 CHI conference on human factors in computing systems}}. \bibinfo{pages}{541--549}.
\newblock


\bibitem[Chevalier et~al\mbox{.}(2022)]%
        {chevalier2022role}
\bibfield{author}{\bibinfo{person}{Morgane Chevalier}, \bibinfo{person}{Christian Giang}, \bibinfo{person}{Laila El-Hamamsy}, \bibinfo{person}{Evgeniia Bonnet}, \bibinfo{person}{Vaios Papaspyros}, \bibinfo{person}{Jean-Philippe Pellet}, \bibinfo{person}{Catherine Audrin}, \bibinfo{person}{Margarida Romero}, \bibinfo{person}{Bernard Baumberger}, {and} \bibinfo{person}{Francesco Mondada}.} \bibinfo{year}{2022}\natexlab{}.
\newblock \showarticletitle{The role of feedback and guidance as intervention methods to foster computational thinking in educational robotics learning activities for primary school}.
\newblock \bibinfo{journal}{\emph{Computers \& Education}}  \bibinfo{volume}{180} (\bibinfo{year}{2022}), \bibinfo{pages}{104431}.
\newblock


\bibitem[Chu et~al\mbox{.}(2025)]%
        {chu2025llm}
\bibfield{author}{\bibinfo{person}{Zhendong Chu}, \bibinfo{person}{Shen Wang}, \bibinfo{person}{Jian Xie}, \bibinfo{person}{Tinghui Zhu}, \bibinfo{person}{Yibo Yan}, \bibinfo{person}{Jinheng Ye}, \bibinfo{person}{Aoxiao Zhong}, \bibinfo{person}{Xuming Hu}, \bibinfo{person}{Jing Liang}, \bibinfo{person}{Philip~S Yu}, {et~al\mbox{.}}} \bibinfo{year}{2025}\natexlab{}.
\newblock \showarticletitle{Llm agents for education: Advances and applications}.
\newblock \bibinfo{journal}{\emph{arXiv preprint arXiv:2503.11733}} (\bibinfo{year}{2025}).
\newblock


\bibitem[Costa and Miranda(2017)]%
        {costa2017relation}
\bibfield{author}{\bibinfo{person}{Joana~M Costa} {and} \bibinfo{person}{Guilhermina~L Miranda}.} \bibinfo{year}{2017}\natexlab{}.
\newblock \showarticletitle{Relation between Alice software and programming learning: A systematic review of the literature and meta-analysis}.
\newblock \bibinfo{journal}{\emph{British Journal of Educational Technology}} \bibinfo{volume}{48}, \bibinfo{number}{6} (\bibinfo{year}{2017}), \bibinfo{pages}{1464--1474}.
\newblock


\bibitem[Elo and Kyng{\"a}s(2008)]%
        {elo2008qualitative}
\bibfield{author}{\bibinfo{person}{Satu Elo} {and} \bibinfo{person}{Helvi Kyng{\"a}s}.} \bibinfo{year}{2008}\natexlab{}.
\newblock \showarticletitle{The qualitative content analysis process}.
\newblock \bibinfo{journal}{\emph{Journal of advanced nursing}} \bibinfo{volume}{62}, \bibinfo{number}{1} (\bibinfo{year}{2008}), \bibinfo{pages}{107--115}.
\newblock


\bibitem[Hoover et~al\mbox{.}(2016)]%
        {hoover2016assessing}
\bibfield{author}{\bibinfo{person}{Amy~K Hoover}, \bibinfo{person}{Jackie Barnes}, \bibinfo{person}{Borna Fatehi}, \bibinfo{person}{Jes{\'u}s Moreno-Le{\'o}n}, \bibinfo{person}{Gillian Puttick}, \bibinfo{person}{Eli Tucker-Raymond}, {and} \bibinfo{person}{Casper Harteveld}.} \bibinfo{year}{2016}\natexlab{}.
\newblock \showarticletitle{Assessing computational thinking in students' game designs}. In \bibinfo{booktitle}{\emph{Proceedings of the 2016 annual symposium on computer-human interaction in play companion extended abstracts}}. \bibinfo{pages}{173--179}.
\newblock


\bibitem[Hou et~al\mbox{.}(2024)]%
        {hou2024codetailor}
\bibfield{author}{\bibinfo{person}{Xinying Hou}, \bibinfo{person}{Zihan Wu}, \bibinfo{person}{Xu Wang}, {and} \bibinfo{person}{Barbara~J Ericson}.} \bibinfo{year}{2024}\natexlab{}.
\newblock \showarticletitle{Codetailor: Llm-powered personalized parsons puzzles for engaging support while learning programming}. In \bibinfo{booktitle}{\emph{Proceedings of the Eleventh ACM Conference on Learning@ Scale}}. \bibinfo{pages}{51--62}.
\newblock


\bibitem[Hsu(2025)]%
        {hsu2025programming}
\bibfield{author}{\bibinfo{person}{Hsiao-Ping Hsu}.} \bibinfo{year}{2025}\natexlab{}.
\newblock \showarticletitle{From programming to prompting: Developing computational thinking through large language model-based generative artificial intelligence}.
\newblock \bibinfo{journal}{\emph{TechTrends}} (\bibinfo{year}{2025}), \bibinfo{pages}{1--22}.
\newblock


\bibitem[Hsu et~al\mbox{.}(2018)]%
        {hsu2018learn}
\bibfield{author}{\bibinfo{person}{Ting-Chia Hsu}, \bibinfo{person}{Shao-Chen Chang}, {and} \bibinfo{person}{Yu-Ting Hung}.} \bibinfo{year}{2018}\natexlab{}.
\newblock \showarticletitle{How to learn and how to teach computational thinking: Suggestions based on a review of the literature}.
\newblock \bibinfo{journal}{\emph{Computers \& Education}}  \bibinfo{volume}{126} (\bibinfo{year}{2018}), \bibinfo{pages}{296--310}.
\newblock


\bibitem[Jiang et~al\mbox{.}(2024)]%
        {jiang2024survey}
\bibfield{author}{\bibinfo{person}{Xuhui Jiang}, \bibinfo{person}{Yuxing Tian}, \bibinfo{person}{Fengrui Hua}, \bibinfo{person}{Chengjin Xu}, \bibinfo{person}{Yuanzhuo Wang}, {and} \bibinfo{person}{Jian Guo}.} \bibinfo{year}{2024}\natexlab{}.
\newblock \showarticletitle{A survey on large language model hallucination via a creativity perspective}.
\newblock \bibinfo{journal}{\emph{arXiv preprint arXiv:2402.06647}} (\bibinfo{year}{2024}).
\newblock


\bibitem[Kalelio{\u{g}}lu(2015)]%
        {kaleliouglu2015new}
\bibfield{author}{\bibinfo{person}{Filiz Kalelio{\u{g}}lu}.} \bibinfo{year}{2015}\natexlab{}.
\newblock \showarticletitle{A new way of teaching programming skills to K-12 students: Code. org}.
\newblock \bibinfo{journal}{\emph{Computers in Human Behavior}}  \bibinfo{volume}{52} (\bibinfo{year}{2015}), \bibinfo{pages}{200--210}.
\newblock


\bibitem[Kasneci et~al\mbox{.}(2023)]%
        {kasneci2023chatgpt}
\bibfield{author}{\bibinfo{person}{Enkelejda Kasneci}, \bibinfo{person}{Kathrin Se{\ss}ler}, \bibinfo{person}{Stefan K{\"u}chemann}, \bibinfo{person}{Maria Bannert}, \bibinfo{person}{Daryna Dementieva}, \bibinfo{person}{Frank Fischer}, \bibinfo{person}{Urs Gasser}, \bibinfo{person}{Georg Groh}, \bibinfo{person}{Stephan G{\"u}nnemann}, \bibinfo{person}{Eyke H{\"u}llermeier}, {et~al\mbox{.}}} \bibinfo{year}{2023}\natexlab{}.
\newblock \showarticletitle{ChatGPT for good? On opportunities and challenges of large language models for education}.
\newblock \bibinfo{journal}{\emph{Learning and individual differences}}  \bibinfo{volume}{103} (\bibinfo{year}{2023}), \bibinfo{pages}{102274}.
\newblock


\bibitem[Kazemitabaar et~al\mbox{.}(2023)]%
        {kazemitabaar2023novices}
\bibfield{author}{\bibinfo{person}{Majeed Kazemitabaar}, \bibinfo{person}{Xinying Hou}, \bibinfo{person}{Austin Henley}, \bibinfo{person}{Barbara~Jane Ericson}, \bibinfo{person}{David Weintrop}, {and} \bibinfo{person}{Tovi Grossman}.} \bibinfo{year}{2023}\natexlab{}.
\newblock \showarticletitle{How novices use LLM-based code generators to solve CS1 coding tasks in a self-paced learning environment}. In \bibinfo{booktitle}{\emph{Proceedings of the 23rd Koli calling international conference on computing education research}}. \bibinfo{pages}{1--12}.
\newblock


\bibitem[Kazimoglu(2020)]%
        {kazimoglu2020enhancing}
\bibfield{author}{\bibinfo{person}{Cagin Kazimoglu}.} \bibinfo{year}{2020}\natexlab{}.
\newblock \showarticletitle{Enhancing confidence in using computational thinking skills via playing a serious game: A case study to increase motivation in learning computer programming}.
\newblock \bibinfo{journal}{\emph{IEEE Access}}  \bibinfo{volume}{8} (\bibinfo{year}{2020}), \bibinfo{pages}{221831--221851}.
\newblock


\bibitem[Leinonen et~al\mbox{.}(2023)]%
        {leinonen2023using}
\bibfield{author}{\bibinfo{person}{Juho Leinonen}, \bibinfo{person}{Arto Hellas}, \bibinfo{person}{Sami Sarsa}, \bibinfo{person}{Brent Reeves}, \bibinfo{person}{Paul Denny}, \bibinfo{person}{James Prather}, {and} \bibinfo{person}{Brett~A Becker}.} \bibinfo{year}{2023}\natexlab{}.
\newblock \showarticletitle{Using large language models to enhance programming error messages}. In \bibinfo{booktitle}{\emph{Proceedings of the 54th ACM Technical Symposium on Computer Science Education V. 1}}. \bibinfo{pages}{563--569}.
\newblock


\bibitem[Lye and Koh(2014)]%
        {lye2014review}
\bibfield{author}{\bibinfo{person}{Sze~Yee Lye} {and} \bibinfo{person}{Joyce Hwee~Ling Koh}.} \bibinfo{year}{2014}\natexlab{}.
\newblock \showarticletitle{Review on teaching and learning of computational thinking through programming: What is next for K-12?}
\newblock \bibinfo{journal}{\emph{Computers in human behavior}}  \bibinfo{volume}{41} (\bibinfo{year}{2014}), \bibinfo{pages}{51--61}.
\newblock


\bibitem[Ma et~al\mbox{.}(2025)]%
        {ma2025dbox}
\bibfield{author}{\bibinfo{person}{Shuai Ma}, \bibinfo{person}{Junling Wang}, \bibinfo{person}{Yuanhao Zhang}, \bibinfo{person}{Xiaojuan Ma}, {and} \bibinfo{person}{April~Yi Wang}.} \bibinfo{year}{2025}\natexlab{}.
\newblock \showarticletitle{Dbox: Scaffolding algorithmic programming learning through learner-llm co-decomposition}. In \bibinfo{booktitle}{\emph{Proceedings of the 2025 CHI Conference on Human Factors in Computing Systems}}. \bibinfo{pages}{1--20}.
\newblock


\bibitem[Mladenovi{\'c} and Hansen(1997)]%
        {mladenovic1997variable}
\bibfield{author}{\bibinfo{person}{Nenad Mladenovi{\'c}} {and} \bibinfo{person}{Pierre Hansen}.} \bibinfo{year}{1997}\natexlab{}.
\newblock \showarticletitle{Variable neighborhood search}.
\newblock \bibinfo{journal}{\emph{Computers \& operations research}} \bibinfo{volume}{24}, \bibinfo{number}{11} (\bibinfo{year}{1997}), \bibinfo{pages}{1097--1100}.
\newblock


\bibitem[Moreno-Le{\'o}n and Robles(2016)]%
        {moreno2016code}
\bibfield{author}{\bibinfo{person}{Jes{\'u}s Moreno-Le{\'o}n} {and} \bibinfo{person}{Gregorio Robles}.} \bibinfo{year}{2016}\natexlab{}.
\newblock \showarticletitle{Code to learn with Scratch? A systematic literature review}. In \bibinfo{booktitle}{\emph{2016 IEEE Global Engineering Education Conference (EDUCON)}}. IEEE, \bibinfo{pages}{150--156}.
\newblock


\bibitem[Nesbitt et~al\mbox{.}(2024)]%
        {nesbitt2024semantic}
\bibfield{author}{\bibinfo{person}{David Nesbitt}, \bibinfo{person}{Dominic Fleischer}, \bibinfo{person}{Edward Callahan}, \bibinfo{person}{Michael Rozenfeld}, {and} \bibinfo{person}{Matthew Taleb}.} \bibinfo{year}{2024}\natexlab{}.
\newblock \bibinfo{title}{Semantic contextual embedding for domain-specific task transfer in large language models}.
\newblock


\bibitem[Pirzado et~al\mbox{.}(2024)]%
        {pirzado2024navigating}
\bibfield{author}{\bibinfo{person}{Farman~Ali Pirzado}, \bibinfo{person}{Awais Ahmed}, \bibinfo{person}{Rom{\'a}n~Alejandro Mendoza-Urdiales}, {and} \bibinfo{person}{Hugo Terashima-Marin}.} \bibinfo{year}{2024}\natexlab{}.
\newblock \showarticletitle{Navigating the pitfalls: Analyzing the behavior of LLMs as a coding assistant for computer science students—a systematic review of the literature}.
\newblock \bibinfo{journal}{\emph{IEEE Access}}  \bibinfo{volume}{12} (\bibinfo{year}{2024}), \bibinfo{pages}{112605--112625}.
\newblock


\bibitem[Prather et~al\mbox{.}(2023)]%
        {prather2023s}
\bibfield{author}{\bibinfo{person}{James Prather}, \bibinfo{person}{Brent~N Reeves}, \bibinfo{person}{Paul Denny}, \bibinfo{person}{Brett~A Becker}, \bibinfo{person}{Juho Leinonen}, \bibinfo{person}{Andrew Luxton-Reilly}, \bibinfo{person}{Garrett Powell}, \bibinfo{person}{James Finnie-Ansley}, {and} \bibinfo{person}{Eddie~Antonio Santos}.} \bibinfo{year}{2023}\natexlab{}.
\newblock \showarticletitle{“It’s weird that it knows what i want”: Usability and interactions with copilot for novice programmers}.
\newblock \bibinfo{journal}{\emph{ACM transactions on computer-human interaction}} \bibinfo{volume}{31}, \bibinfo{number}{1} (\bibinfo{year}{2023}), \bibinfo{pages}{1--31}.
\newblock


\bibitem[Raihan et~al\mbox{.}(2025)]%
        {raihan2025large}
\bibfield{author}{\bibinfo{person}{Nishat Raihan}, \bibinfo{person}{Mohammed~Latif Siddiq}, \bibinfo{person}{Joanna~CS Santos}, {and} \bibinfo{person}{Marcos Zampieri}.} \bibinfo{year}{2025}\natexlab{}.
\newblock \showarticletitle{Large language models in computer science education: A systematic literature review}. In \bibinfo{booktitle}{\emph{Proceedings of the 56th ACM Technical Symposium on Computer Science Education V. 1}}. \bibinfo{pages}{938--944}.
\newblock


\bibitem[Sanford and Naidu(2016)]%
        {sanford2016computational}
\bibfield{author}{\bibinfo{person}{John~F Sanford} {and} \bibinfo{person}{Jaideep~T Naidu}.} \bibinfo{year}{2016}\natexlab{}.
\newblock \showarticletitle{Computational thinking concepts for grade school.}
\newblock \bibinfo{journal}{\emph{Contemporary Issues in Education Research}} \bibinfo{volume}{9}, \bibinfo{number}{1} (\bibinfo{year}{2016}), \bibinfo{pages}{23--32}.
\newblock


\bibitem[Selby and Woollard(2013)]%
        {selby2013computational}
\bibfield{author}{\bibinfo{person}{Cynthia Selby} {and} \bibinfo{person}{John Woollard}.} \bibinfo{year}{2013}\natexlab{}.
\newblock \showarticletitle{Computational thinking: the developing definition}.
\newblock  (\bibinfo{year}{2013}).
\newblock


\bibitem[Shute et~al\mbox{.}(2017)]%
        {shute2017demystifying}
\bibfield{author}{\bibinfo{person}{Valerie~J Shute}, \bibinfo{person}{Chen Sun}, {and} \bibinfo{person}{Jodi Asbell-Clarke}.} \bibinfo{year}{2017}\natexlab{}.
\newblock \showarticletitle{Demystifying computational thinking}.
\newblock \bibinfo{journal}{\emph{Educational research review}}  \bibinfo{volume}{22} (\bibinfo{year}{2017}), \bibinfo{pages}{142--158}.
\newblock


\bibitem[Thirunavukarasu et~al\mbox{.}(2023)]%
        {thirunavukarasu2023large}
\bibfield{author}{\bibinfo{person}{Arun~James Thirunavukarasu}, \bibinfo{person}{Darren Shu~Jeng Ting}, \bibinfo{person}{Kabilan Elangovan}, \bibinfo{person}{Laura Gutierrez}, \bibinfo{person}{Ting~Fang Tan}, {and} \bibinfo{person}{Daniel Shu~Wei Ting}.} \bibinfo{year}{2023}\natexlab{}.
\newblock \showarticletitle{Large language models in medicine}.
\newblock \bibinfo{journal}{\emph{Nature medicine}} \bibinfo{volume}{29}, \bibinfo{number}{8} (\bibinfo{year}{2023}), \bibinfo{pages}{1930--1940}.
\newblock


\bibitem[Tikva and Tambouris(2023)]%
        {tikva2023effect}
\bibfield{author}{\bibinfo{person}{Christina Tikva} {and} \bibinfo{person}{Efthimios Tambouris}.} \bibinfo{year}{2023}\natexlab{}.
\newblock \showarticletitle{The effect of scaffolding programming games and attitudes towards programming on the development of Computational Thinking}.
\newblock \bibinfo{journal}{\emph{Education and Information Technologies}} \bibinfo{volume}{28}, \bibinfo{number}{6} (\bibinfo{year}{2023}), \bibinfo{pages}{6845--6867}.
\newblock


\bibitem[Triantafyllou et~al\mbox{.}(2024)]%
        {triantafyllou2024gamification}
\bibfield{author}{\bibinfo{person}{Serafeim~A Triantafyllou}, \bibinfo{person}{Theodosios Sapounidis}, {and} \bibinfo{person}{Yousef Farhaoui}.} \bibinfo{year}{2024}\natexlab{}.
\newblock \showarticletitle{Gamification and Computational Thinking in Education: A systematic literature review}.
\newblock \bibinfo{journal}{\emph{Salud, Ciencia y Tecnologia-Serie de Conferencias}} \bibinfo{number}{3} (\bibinfo{year}{2024}), \bibinfo{pages}{659}.
\newblock


\bibitem[Troiano et~al\mbox{.}(2019)]%
        {troiano2019my}
\bibfield{author}{\bibinfo{person}{Giovanni~Maria Troiano}, \bibinfo{person}{Sam Snodgrass}, \bibinfo{person}{Erin{\c{c}} Arg{\i}mak}, \bibinfo{person}{Gregorio Robles}, \bibinfo{person}{Gillian Smith}, \bibinfo{person}{Michael Cassidy}, \bibinfo{person}{Eli Tucker-Raymond}, \bibinfo{person}{Gillian Puttick}, {and} \bibinfo{person}{Casper Harteveld}.} \bibinfo{year}{2019}\natexlab{}.
\newblock \showarticletitle{Is my game OK Dr. Scratch? Exploring programming and computational thinking development via metrics in student-designed serious games for STEM}. In \bibinfo{booktitle}{\emph{Proceedings of the 18th ACM international conference on interaction design and children}}. \bibinfo{pages}{208--219}.
\newblock


\bibitem[Vaismoradi et~al\mbox{.}(2013)]%
        {vaismoradi2013content}
\bibfield{author}{\bibinfo{person}{Mojtaba Vaismoradi}, \bibinfo{person}{Hannele Turunen}, {and} \bibinfo{person}{Terese Bondas}.} \bibinfo{year}{2013}\natexlab{}.
\newblock \showarticletitle{Content analysis and thematic analysis: Implications for conducting a qualitative descriptive study}.
\newblock \bibinfo{journal}{\emph{Nursing \& health sciences}} \bibinfo{volume}{15}, \bibinfo{number}{3} (\bibinfo{year}{2013}), \bibinfo{pages}{398--405}.
\newblock


\bibitem[Wang(2025)]%
        {wang2025scaffolding}
\bibfield{author}{\bibinfo{person}{Nicole~C Wang}.} \bibinfo{year}{2025}\natexlab{}.
\newblock \showarticletitle{Scaffolding Creativity: Integrating Generative AI Tools and Real-world Experiences in Business Education}. In \bibinfo{booktitle}{\emph{Proceedings of the Extended Abstracts of the CHI Conference on Human Factors in Computing Systems}}. \bibinfo{pages}{1--9}.
\newblock


\bibitem[Weintrop et~al\mbox{.}(2018)]%
        {weintrop2018starting}
\bibfield{author}{\bibinfo{person}{David Weintrop}, \bibinfo{person}{Alexandria~K Hansen}, \bibinfo{person}{Danielle~B Harlow}, {and} \bibinfo{person}{Diana Franklin}.} \bibinfo{year}{2018}\natexlab{}.
\newblock \showarticletitle{Starting from Scratch: Outcomes of early computer science learning experiences and implications for what comes next}. In \bibinfo{booktitle}{\emph{Proceedings of the 2018 ACM conference on international computing education research}}. \bibinfo{pages}{142--150}.
\newblock


\bibitem[Welz and Lanquillon(2024)]%
        {welz2024enhancing}
\bibfield{author}{\bibinfo{person}{Laslo Welz} {and} \bibinfo{person}{Carsten Lanquillon}.} \bibinfo{year}{2024}\natexlab{}.
\newblock \showarticletitle{Enhancing large language models through external domain knowledge}. In \bibinfo{booktitle}{\emph{International Conference on Human-Computer Interaction}}. Springer, \bibinfo{pages}{135--146}.
\newblock


\bibitem[Wing(2017)]%
        {wing2017computational}
\bibfield{author}{\bibinfo{person}{Jeannette Wing}.} \bibinfo{year}{2017}\natexlab{}.
\newblock \showarticletitle{Computational thinking’s influence on research and education for all}.
\newblock \bibinfo{journal}{\emph{Italian journal of educational technology}} \bibinfo{volume}{25}, \bibinfo{number}{2} (\bibinfo{year}{2017}), \bibinfo{pages}{7--14}.
\newblock


\bibitem[Wing(2006)]%
        {wing2006computational}
\bibfield{author}{\bibinfo{person}{Jeannette~M Wing}.} \bibinfo{year}{2006}\natexlab{}.
\newblock \showarticletitle{Computational thinking}.
\newblock \bibinfo{journal}{\emph{Commun. ACM}} \bibinfo{volume}{49}, \bibinfo{number}{3} (\bibinfo{year}{2006}), \bibinfo{pages}{33--35}.
\newblock


\bibitem[Wood et~al\mbox{.}(1976)]%
        {wood1976role}
\bibfield{author}{\bibinfo{person}{David Wood}, \bibinfo{person}{Jerome~S Bruner}, {and} \bibinfo{person}{Gail Ross}.} \bibinfo{year}{1976}\natexlab{}.
\newblock \showarticletitle{The role of tutoring in problem solving}.
\newblock \bibinfo{journal}{\emph{Journal of child psychology and psychiatry}} \bibinfo{volume}{17}, \bibinfo{number}{2} (\bibinfo{year}{1976}), \bibinfo{pages}{89--100}.
\newblock


\bibitem[Wu et~al\mbox{.}(2025)]%
        {wu2025integrating}
\bibfield{author}{\bibinfo{person}{Chih-Hung Wu}, \bibinfo{person}{Yu-Cheng Chien}, \bibinfo{person}{Mei-Tzu Chou}, {and} \bibinfo{person}{Yueh-Min Huang}.} \bibinfo{year}{2025}\natexlab{}.
\newblock \showarticletitle{Integrating computational thinking, game design, and design thinking: a scoping review on trends, applications, and implications for education}.
\newblock \bibinfo{journal}{\emph{Humanities and Social Sciences Communications}} \bibinfo{volume}{12}, \bibinfo{number}{1} (\bibinfo{year}{2025}), \bibinfo{pages}{1--12}.
\newblock


\bibitem[Yan et~al\mbox{.}(2025)]%
        {yan2025llm}
\bibfield{author}{\bibinfo{person}{Yi-Miao Yan}, \bibinfo{person}{Chuang-Qi Chen}, \bibinfo{person}{Yang-Bang Hu}, {and} \bibinfo{person}{Xin-Dong Ye}.} \bibinfo{year}{2025}\natexlab{}.
\newblock \showarticletitle{LLM-based collaborative programming: impact on students’ computational thinking and self-efficacy}.
\newblock \bibinfo{journal}{\emph{Humanities and Social Sciences Communications}} \bibinfo{volume}{12}, \bibinfo{number}{1} (\bibinfo{year}{2025}), \bibinfo{pages}{1--12}.
\newblock


\bibitem[Zhu et~al\mbox{.}(2025)]%
        {zhu2025exploring}
\bibfield{author}{\bibinfo{person}{Gaoxia Zhu}, \bibinfo{person}{Jason~Fok Kow}, \bibinfo{person}{Xiuyi Fan}, \bibinfo{person}{Ibrahim~H Yeter}, \bibinfo{person}{Lin~Su Chit}, {and} \bibinfo{person}{Yew~Soon Ong}.} \bibinfo{year}{2025}\natexlab{}.
\newblock \showarticletitle{Exploring Undergraduate Students' Computational Thinking Skills Across Engineering Design Processes}.
\newblock \bibinfo{journal}{\emph{Computer Applications in Engineering Education}} \bibinfo{volume}{33}, \bibinfo{number}{3} (\bibinfo{year}{2025}), \bibinfo{pages}{e70035}.
\newblock


\bibitem[Zhu et~al\mbox{.}(2024)]%
        {zhu2024human}
\bibfield{author}{\bibinfo{person}{Gaoxia Zhu}, \bibinfo{person}{Vidya Sudarshan}, \bibinfo{person}{Jason~Fok Kow}, {and} \bibinfo{person}{Yew~Soon Ong}.} \bibinfo{year}{2024}\natexlab{}.
\newblock \showarticletitle{Human-generative AI collaborative problem solving who leads and how students perceive the interactions}. In \bibinfo{booktitle}{\emph{2024 IEEE Conference on Artificial Intelligence (CAI)}}. IEEE, \bibinfo{pages}{680--686}.
\newblock


\end{thebibliography}










\end{document}